\documentclass[prl,twocolumn,showpacs,preprintnumbers,floatfix,amsmath,amssymb,superscriptaddress]{revtex4-1}
\usepackage{color} 

\definecolor{green}{rgb}{0,.5,0}

\definecolor{red}{rgb}{1,0,0}

\usepackage{graphicx}
\usepackage[subfigure]{graphfig}
\usepackage{epsfig}
\usepackage{dcolumn}
\usepackage{amsmath}
\usepackage{multirow}
\usepackage{booktabs}
\usepackage{diagbox}
\usepackage{slashed}
\newcolumntype{d}[1]{D{.}{.}{#1}}

\def\MSbar{\overline{\mathrm{MS}}}

\newcommand{\beq}{\begin{equation}}
\newcommand{\eeq}{\end{equation}}

\immediate\write18{texcount -inc -incbib 
-sum borra.tex > /tmp/wordcount.tex}

\usepackage[colorlinks,linkcolor=blue,citecolor=blue,urlcolor=blue]{hyperref}

\begin{document}

\title{\vspace{1.0in}Total Gluon Helicity Contribution to the Proton Spin from Lattice QCD}

\author{
  \includegraphics[scale=0.25]{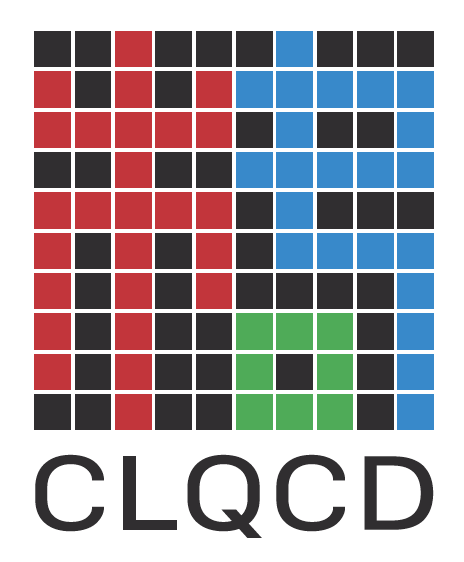}
  \hspace{0.5cm}
  \includegraphics[scale=0.135]{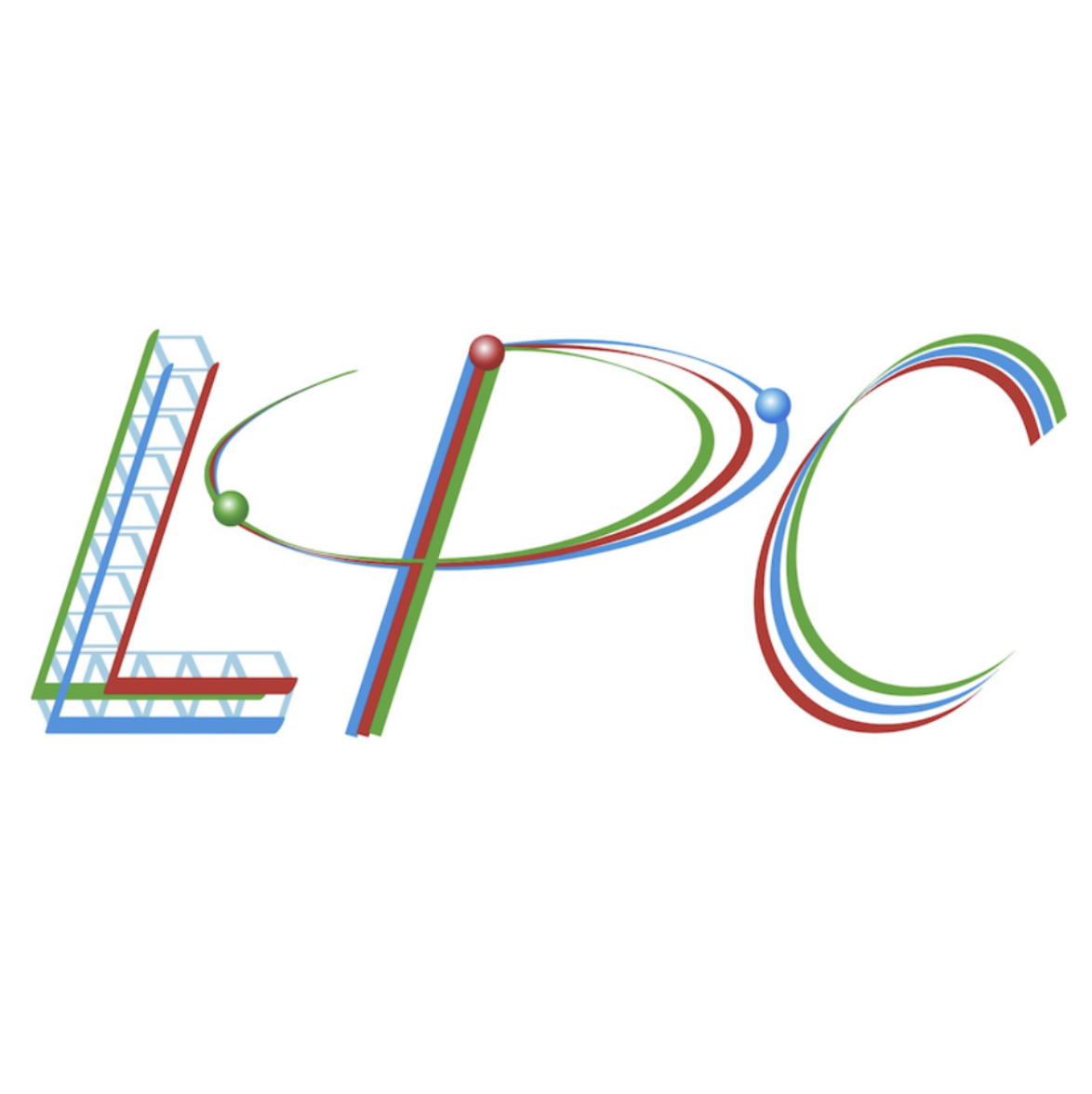}\\
  Dian-Jun Zhao
}
\affiliation{School of Science and Engineering, The Chinese University of Hong Kong, Shenzhen 518172, China}

\author{Long Chen}
\email[Corresponding author: ]{longchen@sdu.edu.cn}
\affiliation{School of Physics, Shandong University, Jinan, Shandong 250100, China}

\author{Hongxin Dong}
\affiliation{Department of Physics and Institute of Theoretical Physics,
Nanjing Normal University, Nanjing, Jiangsu 210023, China}

\author{Xiangdong Ji}
\affiliation{Department of Physics, University of Maryland, College Park, MD 20850, USA}

\author{Liuming Liu}
\affiliation{Institute of Modern Physics, Chinese Academy of Sciences, Lanzhou, 730000, China}
\affiliation{University of Chinese Academy of Sciences, School of Physical Sciences, Beijing 100049, China}

\author{Zhuoyi Pang}
\affiliation{National Centre for Nuclear Research (NCBJ), 02-093 Warsaw, Poland}
\affiliation{School of Science and Engineering, The Chinese University of Hong Kong, Shenzhen 518172, China}

\author{Andreas Sch{\"a}fer}
\affiliation{Institut f{\"u}r Theoretische Physik, Universit{\"a}t Regensburg, D-93040 Regensburg, Germany}
\affiliation{Department of Physics, National Taiwan University, Taipei, Taiwan 106, China}

\author{Peng Sun}
\affiliation{Institute of Modern Physics, Chinese Academy of Sciences, Lanzhou, 730000, China}

\author{Yi-Bo Yang}
\email[Corresponding author: ]{ybyang@itp.ac.cn}
\affiliation{University of Chinese Academy of Sciences, School of Physical Sciences, Beijing 100049, China}
\affiliation{CAS Key Laboratory of Theoretical Physics, Institute of Theoretical Physics, Chinese Academy of Sciences, Beijing 100190, China}
\affiliation{School of Fundamental Physics and Mathematical Sciences, Hangzhou Institute for Advanced Study, UCAS, Hangzhou 310024, China}
\affiliation{International Centre for Theoretical Physics Asia-Pacific, Beijing/Hangzhou, China}

\author{Jian-Hui Zhang}
\email[Corresponding author: ]{zhangjianhui@cuhk.edu.cn}
\affiliation{School of Science and Engineering, The Chinese University of Hong Kong, Shenzhen 518172, China}

\author{Shiyi Zhong}
\affiliation{School of Science and Engineering, The Chinese University of Hong Kong, Shenzhen 518172, China}

\begin{abstract}

We report a state-of-the-art lattice QCD calculation of the total gluon helicity contribution to the proton spin, $\Delta G$. The calculation is done on ensembles with three different lattice spacings $a=\{0.08, 0.09, 0.11\}$ fm. By employing distillation and momentum smearing for proton external states, we extract the bare matrix elements of the topological current $K^\mu$ using 5-HYP smeared Coulomb gauge fixing configurations. Furthermore, we apply a non-perturbative $\mathrm{RI/MOM}$ renormalization scheme augmented by the Cluster Decomposition Error Reduction (CDER) technique to determine the renormalization constants of $K^\mu$. The results obtained from different components $K^{t,i}$ (with $i$ being the direction of proton momentum or polarization) are consistent with Lorentz covariance within uncertainties. After extrapolating to the continuum limit, $\Delta G$ is found to be $\Delta G = 0.231(17)^{\mathrm{sta.}}(44)^{\mathrm{sym.}}$ at the $\overline{\mathrm{MS}}$ scale ${\mu}^2=10\ \mathrm{GeV}^2$, which constitutes approximately $46(9)\%$ of the proton spin.

\end{abstract}

\maketitle

{\it Introduction:}\label{sec:intro}
Since the European Muon Collaboration (EMC) measured the quark spin contribution to the proton spin using deep inelastic scattering of polarized muons and protons~\cite{1988364}, the origin of proton spin has remained one of the most profound puzzles in modern particle and nuclear physics~\cite{Ji:2020ena}. Our current understanding is that the quark spin contribution accounts for only $\sim30\%$ of the proton spin~\cite{deFlorian:2009vb,Nocera:2014gqa,COMPASS:2015mhb,Alexandrou:2017oeh,Lin:2018obj,Liang:2018pis}. The remaining contributions come from quark orbital angular momentum (OAM), as well as gluon spin and OAM.  
Two popular spin sum rules have been proposed to help understanding the proton spin structure. One is the Jaffe-Manohar sum rule~\cite{Jaffe:1989jz}
which offers a complete decomposition of the proton spin into quark and gluon spin and OAM. It has a clear partonic picture in the infinite momentum frame (IMF) of the proton and lightcone gauge. The other is the frame- and gauge-independent sum rule by Ji~\cite{Ji:1996ek}, where the quark and gluon total angular momentum can be connected to the moments of generalized parton distributions~\cite{Muller:1994ses,Ji:1996ek}. 

Among these contributions, the understanding of gluon spin has posed a significant challenge, mainly due to the absence of a local gauge-invariant gluon spin operator~\cite{Ji:2012gc}. Nevertheless, it can be deduced in polarized scattering experiments from the spin-dependent gluon helicity distribution $\Delta g(x)$, more precisely its lowest moment $\Delta G=\int dx\Delta g(x)$. Note that unlike in the case of the quark distribution, the lowest moment of the gluon helicity distribution is still nonlocal. Only in the lightcone gauge does it reduce to a matrix element of the local operator $\vec E\times \vec A$.
Phenomenological global analyses of experimental data indicate that $\Delta G$ is approximately 0.2 at the $\overline{\mathrm{MS}}$ scale  $\mu^2=10\ \mathrm{GeV}^2$, provided that contributions from the small-$x$ region ($x<0.05$) {with large uncertainties} are excluded~\cite{deFlorian:2009vb,deFlorian:2014yva}. 
On the other hand, recent theoretical developments~\cite{Ji:2013fga,Hatta:2013gta,Ji:2014lra}, particularly within the framework of Large-Momentum Effective Theory (LaMET)~\cite{Ji:2013dva,Ji:2014gla,Ji:2020ect}, have enabled a direct computation of $\Delta G$ using lattice QCD. 
In this approach, instead of computing in the IMF and lightcone gauge which are both inaccessible on the lattice, one computes the large-momentum nucleon matrix element of a static gluon spin operator, $\vec{E} \times \vec{A}$, in a physical gauge that leaves the transverse gauge field intact. Actually, there exists an infinite number of operators that can be used to extract $\Delta G$, all belonging to the same universality class~\cite{Hatta:2013gta}. The static gluon spin matrix element can then be matched to $\Delta G$ through a perturbative factorization or matching relation~\cite{Ji:2013fga}. 
Based on this, the $\rm \chi QCD$ collaboration conducted an exploratory calculation of $\Delta G$ by evaluating the nucleon matrix element of $\vec E\times \vec A$ in the Coulomb gauge, which yields the result $0.251(47)(16)$~\cite{Yang:2016plb}. However, the one-loop lattice perturbative renormalization employed in that study introduces a large correction, indicating the necessity of a proper nonperturbative renormalization -- which was, at the time, very challenging to implement due to the Coulomb gauge fixing. Furthermore, recent studies~\cite{Pang:2024sdl} have revealed that the LaMET factorization for $\Delta G$ is incompatible with that for $\Delta g(x)$.

To circumvent the difficulties mentioned above, two new proposals have been made in Ref.~\cite{Pang:2024sdl}. One is based on using the topological current, and the other employs appropriate static nonlocal gluon operators generating the gluon helicity distribution. Both of them have trivial matching only, and therefore resolve the inconsistency previously encountered in the matching for $\Delta G$ and $\Delta g(x)$. Moreover, their nonperturbative renormalization can be easily implemented. This opens a new avenue for systematic studies of $\Delta G$ from lattice QCD. 

In this work, we present a state-of-the-art lattice calculation of the total gluon helicity contribution to the proton spin, $\Delta G$, using the topological current approach proposed in Ref.~\cite{Pang:2024sdl}. The calculation is done on ensembles at three different lattice spacings $a=\{0.08, 0.09, 0.11\}$ fm, which allow us to perform a continuum extrapolation. Distillation~\cite{HadronSpectrum:2009krc}, momentum smearing~\cite{Bali:2016lva,Egerer:2020hnc}, and cluster decomposition~\cite{Liu:2017man} techniques have been used to improve the signal-to-noise ratio of proton matrix elements and parton Green's functions, respectively. The nonperturbative renormalization has been performed in the $\mathrm{RI/MOM}$ scheme, following the operator mixing pattern given in Ref.~\cite{Pang:2024sdl}. After converting to the $\overline{\rm MS}$ scheme, the result is then evolved to the scale $\mu^2=10\, {\rm GeV}^2$ using the renormalization group (RG) equation with three-loop anomalous dimensions.

{\it Theoretical Framework:}\label{sec:theory}
Following Ref.~\cite{Pang:2024sdl}, the total gluon helicity $\Delta G$ can be extracted by computing the matrix element of the local topological current
\begin{equation}
    \label{eq:tp_current}
    K^\mu(x)=\epsilon^{\mu\nu\rho\sigma}\text{Tr}[A_\nu F_{\rho\sigma}-\frac23ig_0A_\nu A_\rho A_\sigma](x)
\end{equation}
in a longitudinally polarized proton. 
More specifically, the proton matrix element of the Coulomb gauge fixed operator $K^\mu$ ($\mu=t,i$) reduces to $\Delta G$ in the IMF, namely, 
\begin{equation}
    \label{eq:threept}
    \frac{\langle \mathrm{PS}_{\mathrm{Proj.i}}|\int d^3x K^{\mu}_{\mathrm{C.G.}}(x)|\mathrm{PS}_{\mathrm{Proj.i}}\rangle}{2S^{\mu}\langle\mathrm{PS}_{\mathrm{Unproj.}}|\mathrm{PS}_{\mathrm{Unproj.}}\rangle}|_{\mathrm{IMF}}= \Delta G,
\end{equation}
$| \mathrm{PS}_{\mathrm{Unproj.}} \rangle$ denotes the proton external state with a specific momentum but without polarization. In contrast, $| \mathrm{PS}_{\mathrm{Proj. i}} \rangle$ represents the state with specific momentum and a definite polarization direction $i$. The $\mu$ component of the corresponding polarization vector in the $i$th-direction is denoted by $S^{\mu}$. The subscript $\mathrm{C.G.}$ indicates the Coulomb gauge fixing for the topological current.

As shown in Ref.~\cite{Pang:2024sdl}, the ultraviolet (UV) divergence of $K^\mu$ is gauge independent. 
Therefore, we can perform a nonperturbative renormalization of $K^\mu$ in the $\mathrm{RI/MOM}$ scheme~\cite{MARTINELLI199581} where the renormalization factor is typically computed in the Landau gauge.  
$K^\mu$ mixes only with the axial-vector current $J^\mu_5=\bar q\gamma_5\gamma^\mu q$ even under RI/MOM renormalization~\cite{Pang:2024sdl}. 
The renormalization constants can be determined by the following conditions
\begin{eqnarray}
    \label{eq:renorm_ori}
    \begin{pmatrix} 
    K^{\mathrm{tree},g/q} \\ 
    J_5^{\mathrm{tree},g/q} 
    \end{pmatrix}
    =
    \begin{pmatrix} 
    Z_{11}^{\mathrm{RI}} & Z_{12}^{\mathrm{RI}} \\ Z_{21}^{\mathrm{RI}} & Z_{22}^{\mathrm{RI}} 
    \end{pmatrix}
    \begin{pmatrix} 
    K^{\mathrm{lat.},g/q} \\ 
    J_5^{\mathrm{lat.},g/q} 
    \end{pmatrix},
\end{eqnarray}
where we introduce the abbreviations $\mathcal{O}^{\mathrm{tree/lat.},{g/q}}\equiv\langle {g/q}|\mathcal{O}|{g/q}\rangle^{\mathrm{tree/lat.}}$ to simplify the notation, where $|g\rangle$ or $|q\rangle$ denotes either gluon or quark external states with specific momentum and polarization.

Note that $K^{\mathrm{tree},q}=J_5^{\mathrm{tree},g}=0$. Moreover, since only disconnected quark loops are involved, we have $J_5^{\mathrm{lat.},g}\approx0$. 
Thus, the RI/MOM renormalization constants can be determined as:
\begin{eqnarray}
    \label{eq:renorm_Z}
    Z_{11}^{\mathrm{RI}}(\mu_{\rm RI}^2)&=&
    \frac{K^\mathrm{tree,g}}{K^{\mathrm{lat.,g}}},\quad\quad
    Z_{12}^{\mathrm{RI}}(\mu_{\rm RI}^2)=-
    \frac{K^{\mathrm{lat.,q}}}
    {J_5^{\mathrm{lat.,q}}}Z_{11}^{\mathrm{RI}},\nonumber\\
    Z_{21}^{\mathrm{RI}}(\mu_{\rm RI}^2)&=&0,\quad\quad\quad\quad\quad
    Z_{22}^{\mathrm{RI}}(\mu_{\rm RI}^2)=
    \frac{J_5^\mathrm{tree,q}}{J_5^{\mathrm{lat.,q}}},
\end{eqnarray}
where $\mu_{\rm RI}$ denotes the renormalization scale in the RI/MOM scheme. With the above approximation, $Z_{22}^{\mathrm{RI}}$ reduces to the axial-vector current renormalization constant $Z_A^\mathrm{RI}$ in Ref.~\cite{Hu:2023jet}. The ratio $K^{\mathrm{lat.,q}}/J_5^{\mathrm{lat.,q}}$ appearing in the off-diagonal element $Z_{12}^{\mathrm{RI}}$ is of $\mathcal{O}(\alpha_s)$. Thus, the $\mathrm{RI/MOM}$-renormalized gluon and quark helicities are given by:
\begin{eqnarray}
    \label{eq:rimom_hel}
    \Delta G^{\mathrm{RI}}&=&Z_{11}^{\mathrm{RI}}
    \big(\Delta G^\mathrm{B.}-
    \frac{K^{\mathrm{lat.,q}}}{J_5^{\mathrm{lat.,q}}
    }\Delta\Sigma^{\mathrm{B.}}\big),\nonumber\\
    \Delta \Sigma^{\mathrm{RI}}&=&Z_A^\mathrm{RI}\Delta\Sigma^{\mathrm{B.}},
\end{eqnarray}
where for simplicity the scale dependence has been suppressed. $\Delta G^\mathrm{B.}$ and $\Delta\Sigma^\mathrm{B.}$ denote the bare gluon and quark helicities extracted from the bare matrix elements of $K^\mu$ and $J_5^\mu$, respectively.

The RI/MOM renormalized results can then be converted to the $\overline{\mathrm{MS}}$ scheme via~\cite{Pang:2024sdl}
\begin{eqnarray}
    \label{eq:rimom_msbar}
    \Delta G^{\overline{\mathrm{MS}}}&=&R_{11}\Delta G^{\mathrm{RI}}+R_{12}\Delta\Sigma^{\mathrm{RI}},\nonumber\\
    \Delta\Sigma^{\overline{\mathrm{MS}}}&=&R_{21}\Delta G^{\mathrm{RI}}+R_{22}\Delta\Sigma^{\mathrm{RI}},
\end{eqnarray}
where again the scale dependence has been suppressed for simplicity. 
In the present work, we have included the conversion factors up to three-loops. Their detailed expressions are given in the Supplemental Material A~\cite{supplemental}.

{\it Lattice Calculation:}\label{sec:latt}
Our calculation is carried out on the $n_f=$2+1 flavor full QCD ensembles generated by CLQCD using the tadpole improved tree level Symanzik (TITLS) gauge action and the tadpole improved tree level Clover (TITLC) fermion action~\cite{Hu:2023jet,CLQCD:2024yyn}. The ensembles used for the simulation are described in Table~\ref{tab:ensem}. The values of $m_{\pi}$ of all the ensembles are chosen to ensure $m_{\pi}L\ge3.7$, thereby efficiently suppressing finite-volume effects. For more details, we refer the readers to the Supplemental Material B~\cite{supplemental}.

\begin{table}[hbt!]                    
\resizebox{1.0\columnwidth}{!}{
\begin{tabular}{| c | c c | c c | c | } 
\hline
Symbol & $\hat{\beta}$& $a$ (fm) & $\tilde{L}^3\times \tilde{T}$ & $m_{\pi}$ (MeV) & $N_{\mathrm{cfg}}$\\
\hline 
C24P29 & \multirow{2}{*}{6.200} & \multirow{2}{*}{0.10524(05)(62)} & $24^3\times 72$ & 292.3(1.0) & 780\\
C48P23 &  &  & $48^3\times 96$ & 224.1(1.2) & 400\\
\hline
E32P29 & 6.308 & 0.08973(20)(53) & $32^3\times 64$ & 287.3(2.5) & 890\\
\hline
F32P30 & 6.410 & 0.07753(03)(45) & $32^3\times 96$ & 300.4(1.2) & 800\\
\hline
\end{tabular}  
}
\caption{Lattice setup used for the simulation, with gauge coupling $\hat{\beta}=10/(g_0^2u_0^4)$, lattice spacing $a$, dimensionless lattice size $\tilde{L}^3\times \tilde{T}$, corresponding pion mass $m_{\pi}$, and number of configurations $N_{\mathrm{cfg}}$.} 
\label{tab:ensem}
\end{table}

In applying Eq.~\eqref{eq:threept}, we choose the proton momentum and polarization to be along three spatial directions and average over them, in order to enhance statistical precision while maintaining rotational symmetry.
Furthermore, 5-HYP smearing is applied to the topological current throughout this calculation.

\begin{figure*}[pt] 
   \centering
   \begin{tabular}{cc}
       \includegraphics[width=0.45\textwidth]{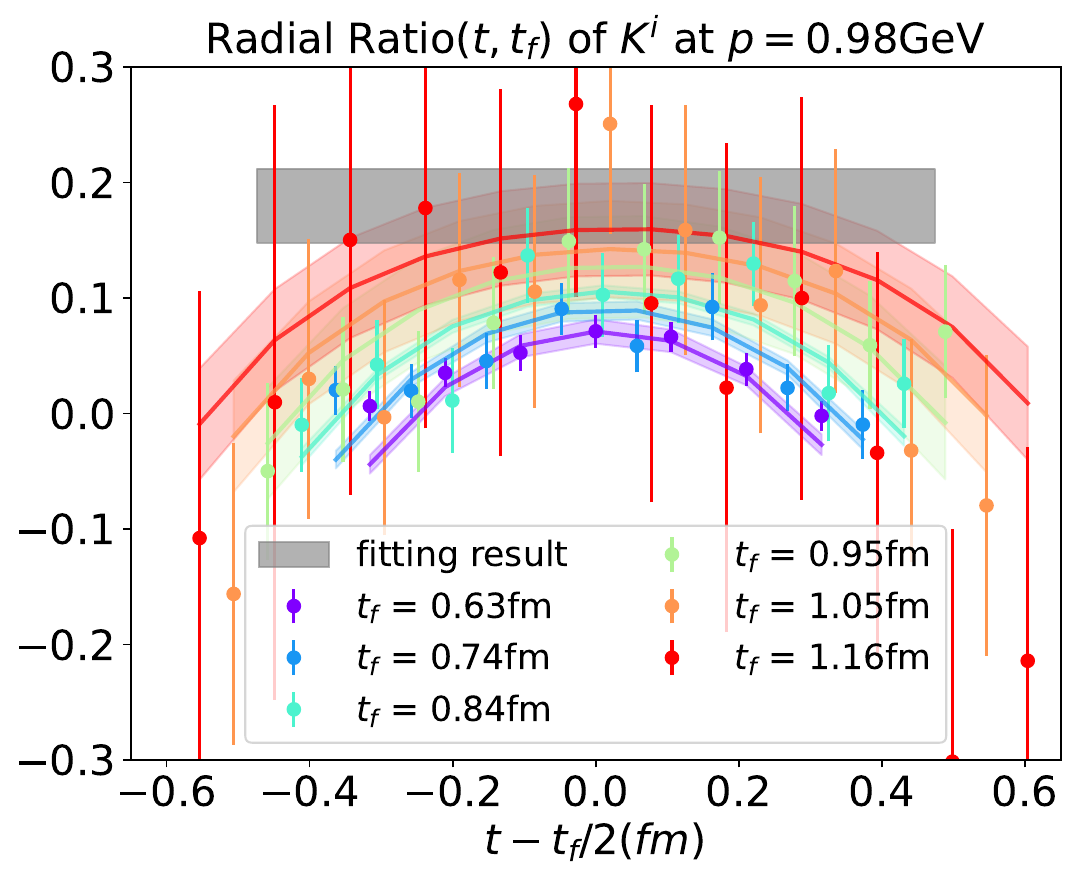}&
       \includegraphics[width=0.45\textwidth]{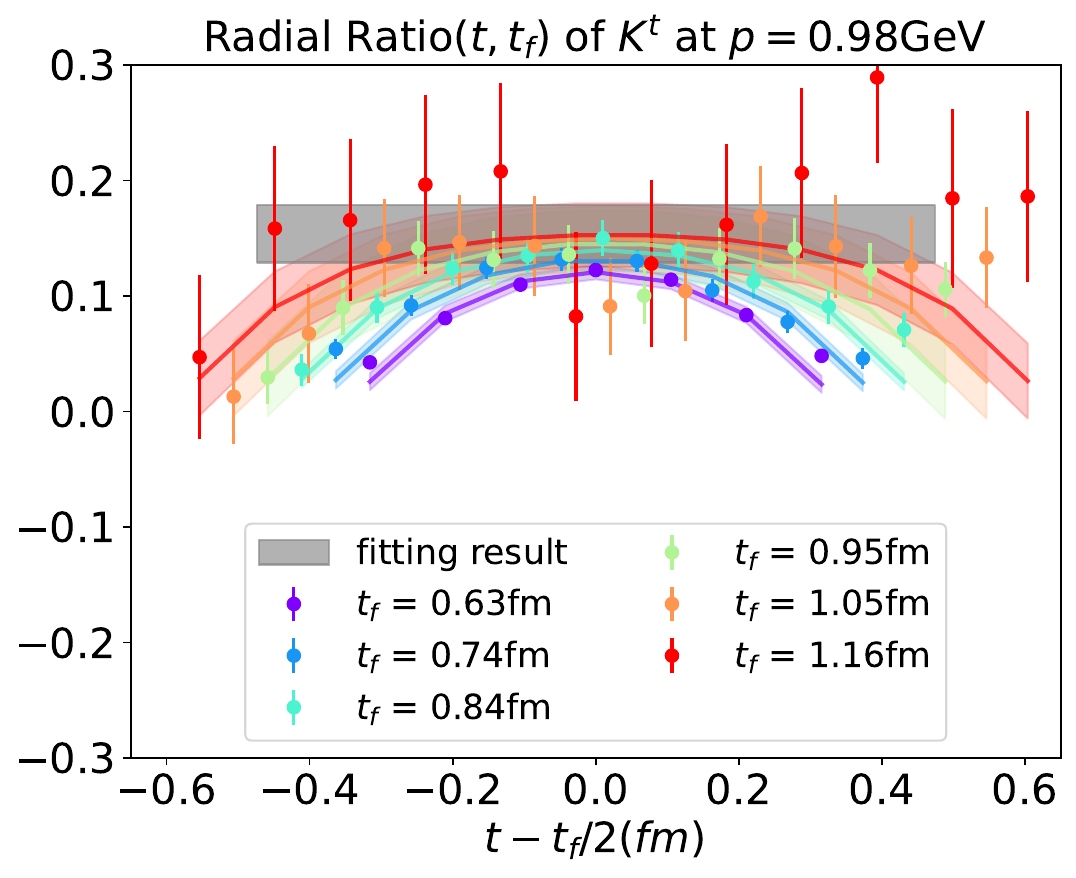} \\
   \end{tabular}
   \caption{The ratio $R_{K^\mu}(t_f, t_i)$ at $t_f\in[0.63,1.16]\ \mathrm{fm}$ and the fitted BME $\langle K^\mu\rangle_N^{\mathrm{B.}}$ (the gray error band) for proton external states with $p=0.98\ \mathrm{GeV}$ and ensemble C24P29, utilizing the fitting procedure described in Eq.~\eqref{eq:ratio_fitting}.}
    \label{fig:kmup2_BME}
\end{figure*}

For proton external states at low momenta ($p\leq 0.5 \ \mathrm{GeV}$), we implement the distillation smearing technique~\cite{HadronSpectrum:2009krc}, while for higher momentum states we employ an approach combining distillation with momentum smearing~\cite{Bali:2016lva,Egerer:2020hnc}.  
We insert the topological current operator into the two-point function of a proton polarized along the $i$th-direction to obtain the disconnected three-point function $C_3(t_f,t_i,t_s)$, where $t_f$, $t_i$, and $t_s$ correspond to sink, current insertion, and source time-slices, respectively. The bare matrix elements (BMEs) of the topological current are extracted through the following ratio:
\begin{equation}
    R_{K^{\mu}}(t_f,t_i,t_s)=\frac{C_3(t_f,t_i,t_s)}{C_2(t_f,t_s)},
\end{equation}
with $C_2(t_f,t_s)$ representing unpolarized proton two-point correlation functions. We sum over all source time-slices: $R_{K^{\mu}}(t_f,t_i)=\sum_{t_s}R_{K^{\mu}}(t_f,t_i,t_s)$, and fit the result to the form
\begin{equation}
\label{eq:ratio_fitting}
    R_{K^{\mu}}(t_f,t_i)=\langle K^\mu\rangle_N^{\mathrm{B.}}+c_1e^{-\Delta E(t_f-t_i)}+c_2e^{-\Delta Et_i},
\end{equation}
where $\langle K^\mu\rangle_N^{\mathrm{B.}}$ is defined as nuclear BME divided by the field operator's normalization factor $2E$, and $c_1,c_2,\Delta E$ are fitting parameters. For illustrative purposes, we show in FIG.~\ref{fig:kmup2_BME} the fitted ground state BME for both $K_i$ and $K_t$ at $p=0.98$ GeV. The fitting results for other momenta can be found in the Supplemental Material B~\cite{supplemental}.

To obtain the renormalization constants, we evaluate the bare Green's function for the topological current
\begin{equation}
    \label{eq:bare_green}
    G_K^{\mu\rho\nu}(p_1,p_2)=\sum_{x,y}e^{-i(p_1\cdot x - p_2\cdot y)}\langle A^\mu(x)K^\rho(0)A^\nu(y)\rangle.
\end{equation}
The bare gluon propagator is given by
\begin{equation}
    \label{eq:gprop}
    S_g^{\mu\nu}(p) =\mathrm{C_V}\langle\mathrm{Tr}[A^\mu(p)A^\nu(-p)]\rangle,
\end{equation}
where $\mathrm{C_V}=2/[(N^2_c-1)\tilde{V}]$, with $N_c=3$ representing the number of colors and $\tilde{V}=\tilde{L}^3\times\tilde{T}$ denoting the dimensionless lattice volume. In addition, we can define the gluon field renormalization constant
\begin{eqnarray}
    \label{eq:Zg}
    Z_{g,\mathrm{diag}}^{\mathrm{RI}}(\mu_{\rm RI}^2)&=&\frac{\sum_{p_\rho=0}1}{\hat p^2\sum_{p_\rho=0}S_{g}^{\rho\rho}(p^2)},
\end{eqnarray}
where $\hat{p}_\mu=\frac{2}{a}\sin{\frac{p_\mu a}{2}}$ represents the lattice momentum, and $\mu_{\rm RI}^2=p^2$ denotes the RI/MOM renormalization scale. $Z_{11}^{\mathrm{RI}}$ in Eq.\eqref{eq:renorm_Z} then takes the following form
\begin{equation}
    \label{eq:Z11}
    Z_{11}^{\mathrm{RI}}(\mu_{\rm RI}^2)=
    \frac{Z_{g,\mathrm{diag}}(\mu_{\rm RI}^2)}{\mathrm{Tr}[S_g^{-1}(p_1)
    G_K(p_1,p_2)
    S_g^{-1}(p_2)(K^{\mathrm{tree,g}})^{-1}]},
\end{equation}
where the external momenta are chosen to satisfy $p_1 = p_2 = p$.
The Green's function of $K^\mathrm{tree,g}$ reads $i\epsilon^{\rho\mu\sigma\nu}\hat{p}_\sigma$ in our normalization convention. We hide the $\mu\rho\nu$ indicator of the bare Green's function $G_K$ and bare gluon propagator $S_g$ in Eq.~\eqref{eq:Z11}. $Z_{11}^{\rm RI}$ then reads, 
\begin{eqnarray}
    \label{eq:Z11_final}
    Z_{11}^{\mathrm{RI}}(\mu_{\rm RI}^2)=\sum_{\mu,\rho,\nu,\sigma}\frac{i\epsilon^{\rho\mu\sigma\nu}\hat{p}_\sigma(S_g^{\mu\mu}+S_g^{\nu\nu})(p^2)}{2\mathrm{C_V}\hat{p}^2
    G_K^{\mu\rho\nu}(p,p)}|_{\mu_{\rm RI}^2=p^2}.
\end{eqnarray}
The summation over indices $\mu,\nu,\rho,\sigma$ in conjunction with the Levi-Civita antisymmetric tensor yields a total of 12 distinct configurations. 
To enhance the signal quality of the pure gluon Green's function $G_K(p,p)$ in Eq.~\eqref{eq:Z11_final}, we abandoned original brute force calculation (which requires huge statistics) and instead adopted the Cluster Decomposition Error Reduction (CDER) technique~\cite{Liu:2017man}, which requires a large volume ensemble.

Due to the limited volume of C24P29, we have chosen to compute the renormalization constants on a larger volume ensemble, C48P23, which shares the same lattice spacing. 
In contrast, for the other two ensembles E32P29 and F32P30, the CDER can be directly applied. For more details on the CDER implementation, we refer readers to the Supplemental Material C~\cite{supplemental}.

\begin{figure}[thb]
\includegraphics[width=0.48\textwidth]{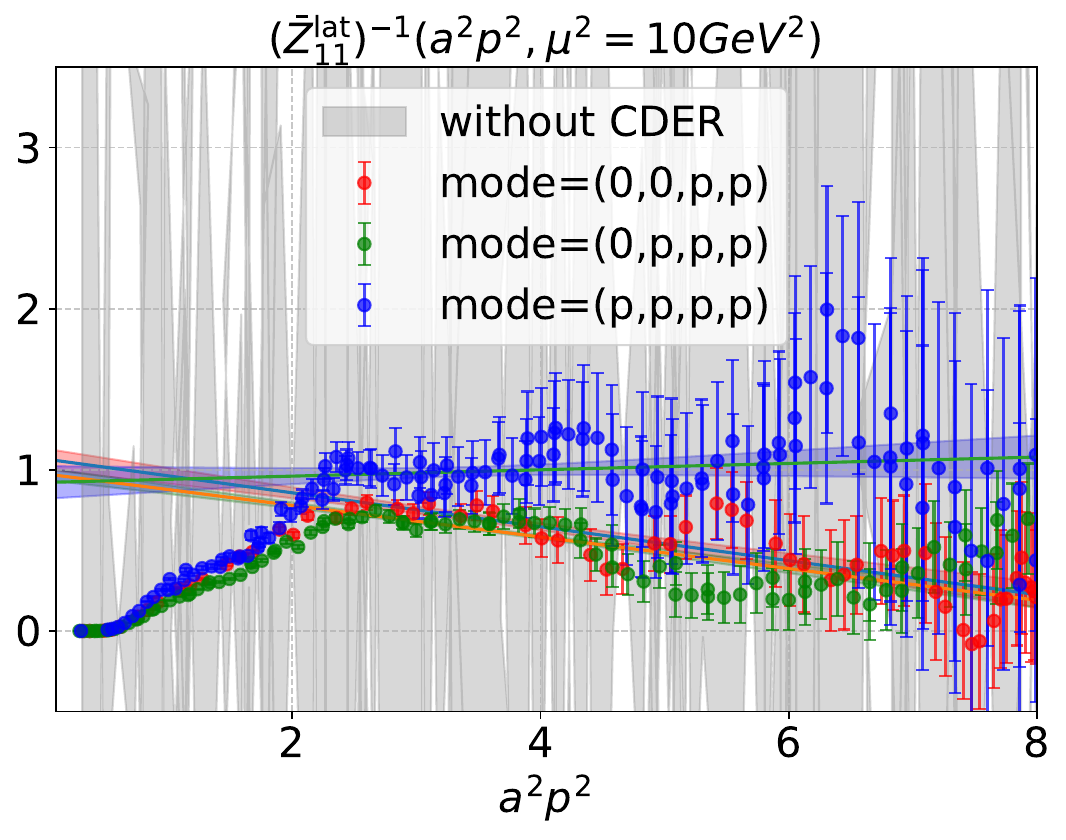}
\caption{
The removal of ${\cal O}(a^2p^2)$ discretization effect in $\bar{Z}_{11}^{\mathrm{lat}}$ for the ensemble C48P23.
The red, green, and blue data points and their error bands represent the CDER results and their $a^2p^2\rightarrow 0$ extrapolated results using momentum modes $(0,0,p,p)$, $(0,p,p,p)$, and $(p,p,p,p)$ in Eq.~\eqref{eq:Z11_final}, where $p=\big(2\pi/(aL)\big)p_{\mathrm{lat}}$ with $p_{\mathrm{lat}}$ represents momentum in the lattice unit.
}
\label{fig:cder_C48P23}
\end{figure}

The conversion factors from the RI/MOM scheme to the $\overline{\rm MS}$ scheme are evaluated by choosing $\mu/\mu_{\rm RI} = e^{-{5}/{6}}$~\cite{Brodsky:1982gc} in the fixed-order expression up to three-loops, and then RG-evolve to $\mu^2=10\,{\rm GeV}^2$. Subsequently, the discretization effects of ${\cal O}(a^2p^2)$ are removed following the strategy proposed in Ref.~\cite{Yang:2018nqn}. Note that here the extrapolation is conducted by varying $p^2$ (and thus $a^2p^2$) and determining the value at $a^2p^2=0$.
To simultaneously ensure the reliability of RI/MOM renormalization and an acceptable signal-to-noise ratio after applying the CDER, we choose the data in the momentum interval $\mu_{\mathrm{RI}}\in[3.0, 5.0]\ \mathrm{GeV}$ for the extrapolation, which correspond to $a^2p^2\in[2.6, 7.1]$ for ensemble C48P23 (the CDER cutoff parameters for $K^\mu$ and $A^\mu$ are $1.26\ \mathrm{fm}$ and $0.74\ \mathrm{fm}$, respectively).
The extrapolated result for the inverse of $\bar{Z}_{11}^{\mathrm{lat}}\equiv R_{11}Z_{11}^{\mathrm{RI}}$ is shown in Fig.~\ref{fig:cder_C48P23}. Similar analyses have also been performed for the other two ensembles with different lattice spacings.

The mixing with quark spin $\Delta\Sigma$, turns out to be an order of magnitude less than $\bar{Z}_{11}^{\mathrm{lat}}$, especially at large $a^2p^2$ (See the Supplemental Material A~\cite{supplemental} for more details).
These properties offer a significant simplification in our numerical analysis, leading us to
\begin{equation}
   \label{eq:MSdelta}
    \Delta G^{\overline{\mathrm{MS}}}\approx \bar{Z}_{11}^{\mathrm{lat}}\Delta G^{\mathrm{B.}},\quad
    \Delta \Sigma^{\overline{\mathrm{MS}}}\approx \bar{Z}_{A}^{\mathrm{lat}}\Delta\Sigma^\mathrm{B.}. 
\end{equation}

\begin{figure}[thb]
\includegraphics[width=0.485\textwidth]{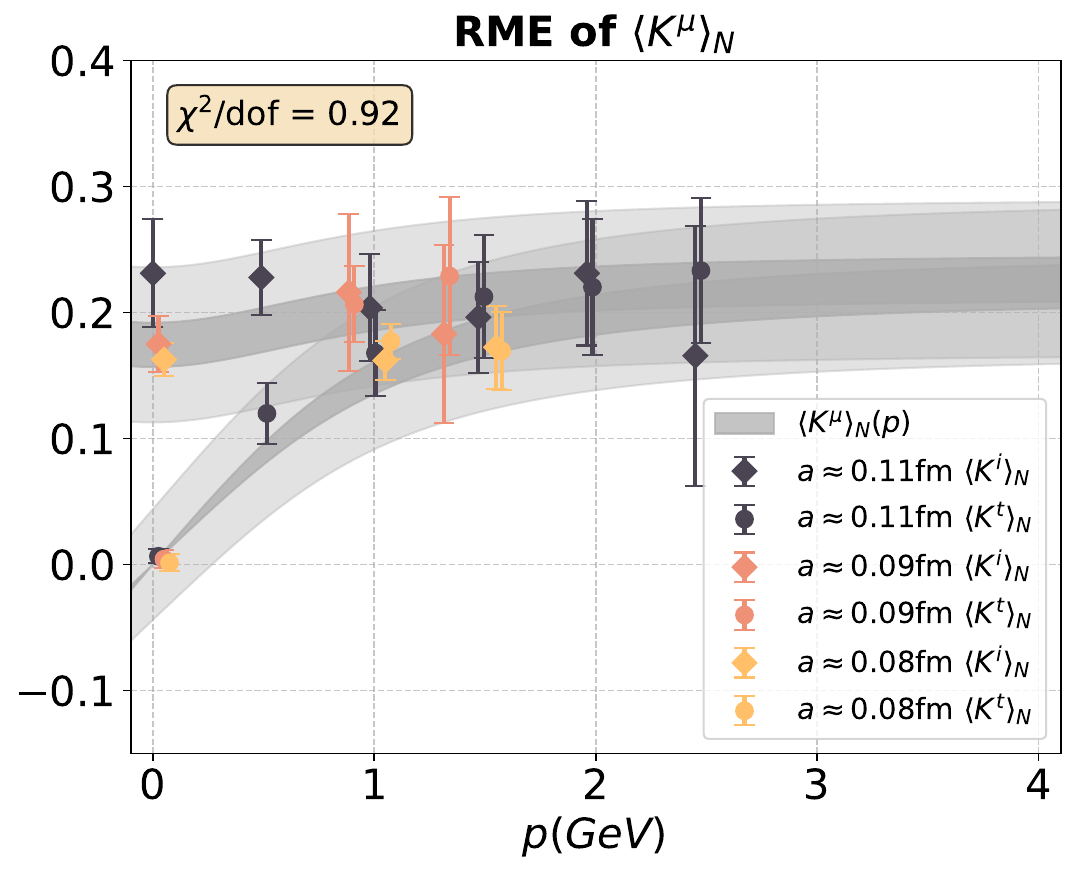}
\caption{The continuum and infinite momentum extrapolation of the total gluon helicity is performed using lattice spacings from $0.08\ \mathrm{fm}$ to $0.11\ \mathrm{fm}$ at the $\overline{\mathrm{MS}}$ scale $\mu^2=10\ \mathrm{GeV}^2$.
The uncertainties are shown as separate bands: statistical uncertainty (dark gray) and all uncertainties that arise during our analysis (light gray).}
\label{fig:continue_extra}
\end{figure}

{\it Numerical Results:}\label{sec:result}
After obtaining the renormalized matrix elements of $K^\mu$ and converting them to the $\overline{\rm MS}$ scheme, we are able to extract the final result by carrying out a continuum and infinite momentum extrapolation. Fig.~\ref{fig:continue_extra} displays $\langle K^\mu\rangle_N\equiv \bar{Z}_{11}^{\mathrm{lat.}}\langle K^\mu\rangle_N^{\mathrm{B.}}$ as a function of lattice spacings and proton momenta. As can be seen from the plot, the results for $\mu=t, i$ exhibit a clear trend consistent with Lorentz covariance within uncertainties. Motivated by this, we adopt the following form for the continuum and infinite momentum extrapolation:
\begin{equation}
    \label{eq:global_fit}
    \langle K^\mu\rangle_N=\frac{S^\mu}{E}\big(\Delta G+\frac{c_{\mathrm{h.t.}}}{E^2}\big) + c_a a^2,
\end{equation}
where $\Delta G$ is the result to be extracted, $c_{\mathrm{h.t.}}$ accounts for higher-twist contributions, and $c_a$ captures discretization effects. 
The pion mass dependence has been neglected in this analysis, primarily due to the significant computational cost of gluon matrix elements at the physical point and the very mild pion mass dependence observed in Ref.~\cite{Yang:2016plb}. 
Since the proton momenta are not significantly larger than its mass, we incorporate the higher-twist contribution using an empirical parametrization form $1/E^2$ in the equation above, where $E = \sqrt{M^2 + p^2}$ with the proton mass $M\approx1\ \mathrm{GeV}$ according to the dispersion relation. 
The primary sources of systematic uncertainty in our analysis are identified as follows: 
1) Including or not the zero-momentum data points in the extrapolation;
2) Discretization errors associated with the finite lattice spacing (the central value difference between the zero momentum $\langle K_i\rangle_N$ for the finest lattice spacing and that extrapolated to the continuum limit); 
3) The choice of the UV window in the RI/MOM scale or $a^2p^2$ (the interval $\mu_{\mathrm{RI}}\in[3.0, 5.0]\ \mathrm{GeV}$ has been extended to $[3.0, 6.0]\ \mathrm{GeV}$ to estimate the corresponding uncertainty);
and 4) Systematic errors associated with the perturbative input (estimated by varying from three-loop to two-loop accuracy). 
As illustrated in FIG.~\ref{fig:continue_extra}, the joint extrapolation yields a gluon helicity contribution $\Delta G^{\overline{\mathrm{MS}},\ 10\ \mathrm{GeV}^2} = 0.231(17)^{\mathrm{sta.}}(44)^{\mathrm{sym.}}$, which constitutes $46(9)\%$ of the proton's total spin.

{\it Summary:}\label{sec:summary}
In this work, we presented a state-of-the-art lattice QCD calculation of the total gluon helicity contribution, $\Delta G$, to the proton spin. The calculation leveraged appropriate components of the topological current operator, $K^\mu$ (specifically $\mu=\{t,i\}$ where $i$ is along the direction of the proton momentum). The bare matrix element is nonperturbatively renormalized in the RI/MOM scheme and subsequently converted to the $\overline{\rm MS}$ scheme via a three-loop conversion factor with a RG improvement. The result is then extrapolated to the continuum and infinite momentum limits, yielding $\Delta G = 0.231(17)^{\mathrm{sta.}}(44)^{\mathrm{sym.}}$ at $\mu^2=10\ \mathrm{GeV}^2$. This corresponds to approximately $46(9)\%$ of the proton spin and clearly rules out the possibility of a negative gluon helicity contribution~\cite{Zhou:2022wzm,Whitehill:2022mpq,Hunt-Smith:2024khs}. The finding offers important insight for future theoretical and experimental efforts aimed at resolving the proton spin puzzle. We have also taken into account various sources of systematic uncertainties, including those from the exclusion of zero-momentum points, discretization errors, and the choice of the renormalization scale. 

Moreover, our analysis demonstrated, for the first time, that the temporal and spatial components of $K^\mu$ generate consistent results for $\Delta G$ when the proton is boosted to the IMF. This is physically intuitive, as these two components are predominantly sensitive to color electric and magnetic fields, respectively, which become of comparable magnitude in the IMF.

Our calculation showcases the capability of combining high-precision extraction of disconnected matrix elements with advanced renormalization techniques to address challenging problems in lattice QCD. The result offers important guidance for future experimental measurements of gluon spin and paves the way for subsequent calculations of quark and gluon OAM, ultimately enabling a complete first-principles decomposition of the proton spin.

\section*{Acknowledgment}
{We thank
Fangcheng He, Zhi-Cheng Hu, Xiangyu Jiang, Li-Jun Zhou, 
and other CLQCD and LPC members 
for valuable comments and suggestions. 
We thank the CLQCD collaborations for providing us their gauge configurations with dynamical fermions~\cite{Hu:2023jet}, which are generated on HPC Cluster of ITP-CAS, IHEP-CAS and CSNS-CAS, the Southern Nuclear Science Computing Center(SNSC) and the Siyuan-1 cluster supported by the Center for High Performance Computing at Shanghai Jiao Tong University. 
The calculations were performed using the PyQUDA~\cite{Jiang:2024lto} and Chroma~\cite{Edwards:2004sx} software suite with QUDA~\cite{Clark:2009wm,Babich:2011np,Clark:2016rdz} through HIP programming model~\cite{Bi:2020wpt}. The numerical calculation were carried out on the ORISE Supercomputer, the Southern Nuclear Science Computing Center (SNSC). This work is supported in part by NSFC grants No. 12375080, 12525504, 12435002, 12293060, 12293062, and 12447101, No.~12205171, No.~12235008, No.~12321005, National Key R\&D Program of China No.2024YFE0109800, the Ministry of Science and Technology of China under Grant No. 2024YFA1611004, by CUHK-Shenzhen under grant No. UDF01002851, the Strategic Priority Research Program of Chinese Academy of Sciences, Grant No.\ YSBR-101, and Department of Science and Technology of Shandong province No.~tsqn202312052 and~2024HWYQ-005.}

\bibliography{reference}

@article{Egerer:2020hnc,
    author = "Egerer, Colin and Edwards, Robert G. and Orginos, Kostas and Richards, David G.",
    title = "{Distillation at High-Momentum}",
    eprint = "2009.10691",
    archivePrefix = "arXiv",
    primaryClass = "hep-lat",
    reportNumber = "JLAB-THY-20-3250",
    doi = "10.1103/PhysRevD.103.034502",
    journal = "Phys. Rev. D",
    volume = "103",
    number = "3",
    pages = "034502",
    year = "2021"
}

@article{Liang:2018pis,
    author = "Liang, Jian and Yang, Yi-Bo and Draper, Terrence and Gong, Ming and Liu, Keh-Fei",
    title = "{Quark spins and Anomalous Ward Identity}",
    eprint = "1806.08366",
    archivePrefix = "arXiv",
    primaryClass = "hep-ph",
    doi = "10.1103/PhysRevD.98.074505",
    journal = "Phys. Rev. D",
    volume = "98",
    number = "7",
    pages = "074505",
    year = "2018"
}

@article{Edwards:2004sx,
	archiveprefix = {arXiv},
	author = {Edwards, Robert G. and Joo, Balint},
	collaboration = {SciDAC, LHPC, UKQCD},
	doi = {10.1016/j.nuclphysbps.2004.11.254},
	editor = {Bodwin, Geoffrey T. and Sinclair, D. K. and Eichten, E. and Holmgren, D. and Kronfeld, Andreas S. and Mackenzie, P. and Okamoto, M. and Simone, J. and El-Khadra, Aida X.},
	eprint = {hep-lat/0409003},
	journal = {Nucl. Phys. B Proc. Suppl.},
	pages = {832},
	reportnumber = {JLAB-THY-04-54},
	title = {{The Chroma software system for lattice QCD}},
	volume = {140},
	year = {2005}}

@article{Bali:2016lva,
    author = {Bali, Gunnar S. and Lang, Bernhard and Musch, Bernhard U. and Sch{\"a}fer, Andreas},
    title = "{Novel quark smearing for hadrons with high momenta in lattice QCD}",
    eprint = "1602.05525",
    archivePrefix = "arXiv",
    primaryClass = "hep-lat",
    doi = "10.1103/PhysRevD.93.094515",
    journal = "Phys. Rev. D",
    volume = "93",
    number = "9",
    pages = "094515",
    year = "2016"
}

@article{Hatta:2013gta,
    author = "Hatta, Yoshitaka and Ji, Xiangdong and Zhao, Yong",
    title = "{Gluon helicity $\Delta G$ from a universality class of operators on a lattice}",
    eprint = "1310.4263",
    archivePrefix = "arXiv",
    primaryClass = "hep-ph",
    reportNumber = "YITP-13-111",
    doi = "10.1103/PhysRevD.89.085030",
    journal = "Phys. Rev. D",
    volume = "89",
    number = "8",
    pages = "085030",
    year = "2014"
}

@article{Yang:2016plb,
    author = "Yang, Yi-Bo and Sufian, Raza Sabbir and Alexandru, Andrei and Draper, Terrence and Glatzmaier, Michael J. and Liu, Keh-Fei and Zhao, Yong",
    title = "{Glue Spin and Helicity in the Proton from Lattice QCD}",
    eprint = "1609.05937",
    archivePrefix = "arXiv",
    primaryClass = "hep-ph",
    doi = "10.1103/PhysRevLett.118.102001",
    journal = "Phys. Rev. Lett.",
    volume = "118",
    number = "10",
    pages = "102001",
    year = "2017"
}

@article{deFlorian:2014yva,
    author = "de Florian, Daniel and Sassot, Rodolfo and Stratmann, Marco and Vogelsang, Werner",
    title = "{Evidence for polarization of gluons in the proton}",
    eprint = "1404.4293",
    archivePrefix = "arXiv",
    primaryClass = "hep-ph",
    doi = "10.1103/PhysRevLett.113.012001",
    journal = "Phys. Rev. Lett.",
    volume = "113",
    number = "1",
    pages = "012001",
    year = "2014"
}

@article{Lin:2018obj,
    author = "Lin, Huey-Wen and Gupta, Rajan and Yoon, Boram and Jang, Yong-Chull and Bhattacharya, Tanmoy",
    title = "{Quark contribution to the proton spin from 2+1+1-flavor lattice QCD}",
    eprint = "1806.10604",
    archivePrefix = "arXiv",
    primaryClass = "hep-lat",
    reportNumber = "LA-UR-18-25337, MSUHEP-18-010",
    doi = "10.1103/PhysRevD.98.094512",
    journal = "Phys. Rev. D",
    volume = "98",
    number = "9",
    pages = "094512",
    year = "2018"
}

@article{Alexandrou:2017oeh,
    author = "Alexandrou, C. and Constantinou, M. and Hadjiyiannakou, K. and Jansen, K. and Kallidonis, C. and Koutsou, G. and Vaquero Avil{\'e}s-Casco, A. and Wiese, C.",
    title = "{Nucleon Spin and Momentum Decomposition Using Lattice QCD Simulations}",
    eprint = "1706.02973",
    archivePrefix = "arXiv",
    primaryClass = "hep-lat",
    reportNumber = "DESY-17-086",
    doi = "10.1103/PhysRevLett.119.142002",
    journal = "Phys. Rev. Lett.",
    volume = "119",
    number = "14",
    pages = "142002",
    year = "2017"
}

@article{Ji:1996ek,
    author = "Ji, Xiang-Dong",
    title = "{Gauge-Invariant Decomposition of Nucleon Spin}",
    eprint = "hep-ph/9603249",
    archivePrefix = "arXiv",
    reportNumber = "MIT-CTP-2517",
    doi = "10.1103/PhysRevLett.78.610",
    journal = "Phys. Rev. Lett.",
    volume = "78",
    pages = "610--613",
    year = "1997"
}

@article{Ji:2013fga,
    author = "Ji, Xiangdong and Zhang, Jian-Hui and Zhao, Yong",
    title = "{Physics of the Gluon-Helicity Contribution to Proton Spin}",
    eprint = "1304.6708",
    archivePrefix = "arXiv",
    primaryClass = "hep-ph",
    reportNumber = "PP-013-004",
    doi = "10.1103/PhysRevLett.111.112002",
    journal = "Phys. Rev. Lett.",
    volume = "111",
    pages = "112002",
    year = "2013"
}

@article{Muller:1994ses,
    author = {M{\"u}ller, Dieter and Robaschik, D. and Geyer, B. and Dittes, F. -M. and Ho{\v{r}}ej{\v{s}}i, J.},
    title = "{Wave functions, evolution equations and evolution kernels from light ray operators of QCD}",
    eprint = "hep-ph/9812448",
    archivePrefix = "arXiv",
    reportNumber = "NTZ-6-91, NTZ-91-6",
    doi = "10.1002/prop.2190420202",
    journal = "Fortsch. Phys.",
    volume = "42",
    pages = "101--141",
    year = "1994"
}

@article{Yang:2018bft,
    author = "Yang, Yi-Bo and Gong, Ming and Liang, Jian and Lin, Huey-Wen and Liu, Keh-Fei and Pefkou, Dimitra and Shanahan, Phiala",
    title = "{Nonperturbatively renormalized glue momentum fraction at the physical pion mass from lattice QCD}",
    eprint = "1805.00531",
    archivePrefix = "arXiv",
    primaryClass = "hep-lat",
    reportNumber = "JLAB-THY-18-2739",
    doi = "10.1103/PhysRevD.98.074506",
    journal = "Phys. Rev. D",
    volume = "98",
    number = "7",
    pages = "074506",
    year = "2018"
}

@article{COMPASS:2015mhb,
    author = "Adolph, C. and others",
    collaboration = "COMPASS",
    title = "{The spin structure function $g_1^{\rm p}$ of the proton and a test of the Bjorken sum rule}",
    eprint = "1503.08935",
    archivePrefix = "arXiv",
    primaryClass = "hep-ex",
    reportNumber = "CERN-PH-EP-2015-085",
    doi = "10.1016/j.physletb.2015.11.064",
    journal = "Phys. Lett. B",
    volume = "753",
    pages = "18--28",
    year = "2016"
}

@article{Nocera:2014gqa,
    author = "Nocera, Emanuele R. and Ball, Richard D. and Forte, Stefano and Ridolfi, Giovanni and Rojo, Juan",
    collaboration = "NNPDF",
    title = "{A first unbiased global determination of polarized PDFs and their uncertainties}",
    eprint = "1406.5539",
    archivePrefix = "arXiv",
    primaryClass = "hep-ph",
    reportNumber = "CERN-PH-TH-2014-106, IFUN-1028-FT, EDINBURGH-14-11, OUTP-14-06P",
    doi = "10.1016/j.nuclphysb.2014.08.008",
    journal = "Nucl. Phys. B",
    volume = "887",
    pages = "276--308",
    year = "2014"
}

@article{deFlorian:2009vb,
    author = "de Florian, Daniel and Sassot, Rodolfo and Stratmann, Marco and Vogelsang, Werner",
    title = "{Extraction of Spin-Dependent Parton Densities and Their Uncertainties}",
    eprint = "0904.3821",
    archivePrefix = "arXiv",
    primaryClass = "hep-ph",
    doi = "10.1103/PhysRevD.80.034030",
    journal = "Phys. Rev. D",
    volume = "80",
    pages = "034030",
    year = "2009"
}

@article{1988364,
title = {A measurement of the spin asymmetry and determination of the structure function g1 in deep inelastic muon-proton scattering},
journal = {Physics Letters B},
volume = {206},
number = {2},
pages = {364-370},
year = {1988},
issn = {0370-2693},
doi = {https://doi.org/10.1016/0370-2693(88)91523-7},
url = {https://www.sciencedirect.com/science/article/pii/0370269388915237},
author = {J. Ashman et al}
}

@article{Ji:2020ena,
    author = "Ji, Xiangdong and Yuan, Feng and Zhao, Yong",
    title = "{What we know and what we don{\textquoteright}t know about the proton spin after 30 years}",
    eprint = "2009.01291",
    archivePrefix = "arXiv",
    primaryClass = "hep-ph",
    doi = "10.1038/s42254-020-00248-4",
    journal = "Nature Rev. Phys.",
    volume = "3",
    number = "1",
    pages = "27--38",
    year = "2021"
}

@article{Ji:2012gc,
    author = "Ji, Xiangdong and Xu, Yang and Zhao, Yong",
    title = "{Gluon Spin, Canonical Momentum, and Gauge Symmetry}",
    eprint = "1205.0156",
    archivePrefix = "arXiv",
    primaryClass = "hep-ph",
    reportNumber = "UMD-DOE-40762-519",
    doi = "10.1007/JHEP08(2012)082",
    journal = "JHEP",
    volume = "08",
    pages = "082",
    year = "2012"
}

@article{Ji:2014lra,
    author = "Ji, Xiangdong and Zhang, Jian-Hui and Zhao, Yong",
    title = "{Justifying the Naive Partonic Sum Rule for Proton Spin}",
    eprint = "1409.6329",
    archivePrefix = "arXiv",
    primaryClass = "hep-ph",
    doi = "10.1016/j.physletb.2015.02.054",
    journal = "Phys. Lett. B",
    volume = "743",
    pages = "180--183",
    year = "2015"
}

@article{Zhou:2022wzm,
    author = "Zhou, Y. and Sato, N. and Melnitchouk, W.",
    collaboration = "Jefferson Lab Angular Momentum (JAM)",
    title = "{How well do we know the gluon polarization in the proton?}",
    eprint = "2201.02075",
    archivePrefix = "arXiv",
    primaryClass = "hep-ph",
    reportNumber = "JLAB-THY-22-3462",
    doi = "10.1103/PhysRevD.105.074022",
    journal = "Phys. Rev. D",
    volume = "105",
    number = "7",
    pages = "074022",
    year = "2022"
}

@article{Whitehill:2022mpq,
    author = "Whitehill, R. M. and Zhou, Yiyu and Sato, N. and Melnitchouk, W.",
    collaboration = "Jefferson Lab Angular Momentum (JAM)",
    title = "{Accessing gluon polarization with high-PT hadrons in SIDIS}",
    eprint = "2210.12295",
    archivePrefix = "arXiv",
    primaryClass = "hep-ph",
    reportNumber = "JLAB-THY-22-3743, ADP-22-30/T1201",
    doi = "10.1103/PhysRevD.107.034033",
    journal = "Phys. Rev. D",
    volume = "107",
    number = "3",
    pages = "034033",
    year = "2023"
}

@article{Hunt-Smith:2024khs,
    author = "Hunt-Smith, N. T. and Cocuzza, C. and Melnitchouk, W. and Sato, N. and Thomas, A. W. and White, M. J.",
    collaboration = "JAM",
    title = "{New Data-Driven Constraints on the Sign of Gluon Polarization in the Proton}",
    eprint = "2403.08117",
    archivePrefix = "arXiv",
    primaryClass = "hep-ph",
    reportNumber = "JLAB-THY-24-4003, ADP-24-04/T1243",
    doi = "10.1103/PhysRevLett.133.161901",
    journal = "Phys. Rev. Lett.",
    volume = "133",
    number = "16",
    pages = "161901",
    year = "2024"
}

@article{MARTINELLI199581,
title = {A general method for non-perturbative renormalization of lattice operators},
journal = {Nuclear Physics B},
volume = {445},
number = {1},
pages = {81-105},
year = {1995},
issn = {0550-3213},
doi = {https://doi.org/10.1016/0550-3213(95)00126-D},
url = {https://www.sciencedirect.com/science/article/pii/055032139500126D},
author = {G. Martinelli and C. Pittori and C.T. Sachrajda and M. Testa and A. Vladikas}
}

@article{Liu:2017man,
    author = "Liu, Keh-Fei and Liang, Jian and Yang, Yi-Bo",
    title = "{Variance Reduction and Cluster Decomposition}",
    eprint = "1705.06358",
    archivePrefix = "arXiv",
    primaryClass = "hep-lat",
    reportNumber = "MSUHEP-17-006",
    doi = "10.1103/PhysRevD.97.034507",
    journal = "Phys. Rev. D",
    volume = "97",
    number = "3",
    pages = "034507",
    year = "2018"
}

@article{Jiang:2024lto,
    author = "Jiang, Xiangyu and Shi, Chunjiang and Chen, Ying and Gong, Ming and Yang, Yi-Bo",
    title = "{Use QUDA for lattice QCD calculation with Python}",
    eprint = "2411.08461",
    archivePrefix = "arXiv",
    primaryClass = "hep-lat",
    month = "11",
    year = "2024"
}

@article{HadronSpectrum:2009krc,
    author = "Peardon, Michael and Bulava, John and Foley, Justin and Morningstar, Colin and Dudek, Jozef and Edwards, Robert G. and Joo, Balint and Lin, Huey-Wen and Richards, David G. and Juge, Keisuke Jimmy",
    collaboration = "Hadron Spectrum",
    title = "{A Novel quark-field creation operator construction for hadronic physics in lattice QCD}",
    eprint = "0905.2160",
    archivePrefix = "arXiv",
    primaryClass = "hep-lat",
    reportNumber = "JLAB-THY-09-985",
    doi = "10.1103/PhysRevD.80.054506",
    journal = "Phys. Rev. D",
    volume = "80",
    pages = "054506",
    year = "2009"
}

@article{Jaffe:1989jz,
    author = "Jaffe, R. L. and Manohar, Aneesh",
    title = "{The $g_1$ Problem: Fact and Fantasy on the Spin of the Proton}",
    reportNumber = "MIT-CTP-1706-REV, MIT-CTP-1706",
    doi = "10.1016/0550-3213(90)90506-9",
    journal = "Nucl. Phys. B",
    volume = "337",
    pages = "509--546",
    year = "1990"
}

@article{Pang:2024sdl,
    author = "Pang, Zhuoyi and Yao, Fei and Zhang, Jian-Hui",
    title = "{Total gluon helicity from lattice without effective theory matching}",
    eprint = "2404.00693",
    archivePrefix = "arXiv",
    primaryClass = "hep-ph",
    doi = "10.1007/JHEP07(2024)222",
    journal = "JHEP",
    volume = "07",
    pages = "222",
    year = "2024"
}

@article{Hu:2023jet,
    author = "Hu, Zhi-Cheng and others",
    collaboration = "CLQCD",
    title = "{Quark masses and low-energy constants in the continuum from the tadpole-improved clover ensembles}",
    eprint = "2310.00814",
    archivePrefix = "arXiv",
    primaryClass = "hep-lat",
    doi = "10.1103/PhysRevD.109.054507",
    journal = "Phys. Rev. D",
    volume = "109",
    number = "5",
    pages = "054507",
    year = "2024"
}

@article{Ji:2014gla,
	archiveprefix = {arXiv},
	author = {Ji, Xiangdong},
	doi = {10.1007/s11433-014-5492-3},
	eprint = {1404.6680},
	journal = {Sci. China Phys. Mech. Astron.},
	pages = {1407-1412},
	primaryclass = {hep-ph},
	slaccitation = {%%CITATION = ARXIV:1404.6680;%%},
	title = {{Parton Physics from Large-Momentum Effective Field Theory}},
	volume = {57},
	year = {2014}}

@article{Bi:2020wpt,
	archiveprefix = {arXiv},
	author = {Bi, Yu-Jiang and Xiao, Yi and Gong, Ming and Guo, Wei-Yi and Sun, Peng and Xu, Shun and Yang, Yi-Bo},
	booktitle = {{Proceedings, 37th International Symposium on Lattice Field Theory (Lattice 2019): Wuhan, China, June 16-22 2019}},
	doi = {10.22323/1.363.0286},
	eprint = {2001.05706},
	journal = {PoS},
	pages = {286},
	primaryclass = {hep-lat},
	slaccitation = {%%CITATION = ARXIV:2001.05706;%%},
	title = {{Lattice QCD package GWU-code and QUDA with HIP}},
	volume = {LATTICE2019},
	year = {2020}}

@article{Clark:2016rdz,
	archiveprefix = {arXiv},
	author = {Clark, M. A. and Jo, Blint and Strelchenko, Alexei and Cheng, Michael and Gambhir, Arjun and Brower, Richard},
	eprint = {1612.07873},
	primaryclass = {hep-lat},
	reportnumber = {FERMILAB-CONF-16-638-CD},
	slaccitation = {%%CITATION = ARXIV:1612.07873;%%},
	title = {{Accelerating Lattice QCD Multigrid on GPUs Using Fine-Grained Parallelization}},
	year = {2016}}

@inproceedings{Babich:2011np,
	archiveprefix = {arXiv},
	author = {Babich, R. and Clark, M. A. and Joo, B. and Shi, G. and Brower, R. C. and Gottlieb, S.},
	booktitle = {{SC11 International Conference for High Performance Computing, Networking, Storage and Analysis Seattle, Washington, November 12-18, 2011}},
	doi = {10.1145/2063384.2063478},
	eprint = {1109.2935},
	primaryclass = {hep-lat},
	slaccitation = {%%CITATION = ARXIV:1109.2935;%%},
	title = {{Scaling Lattice QCD beyond 100 GPUs}},
	year = {2011}}

@article{Clark:2009wm,
	archiveprefix = {arXiv},
	author = {Clark, M. A. and Babich, R. and Barros, K. and Brower, R. C. and Rebbi, C.},
	doi = {10.1016/j.cpc.2010.05.002},
	eprint = {0911.3191},
	journal = {Comput. Phys. Commun.},
	pages = {1517-1528},
	primaryclass = {hep-lat},
	slaccitation = {%%CITATION = ARXIV:0911.3191;%%},
	title = {{Solving Lattice QCD systems of equations using mixed precision solvers on GPUs}},
	volume = {181},
	year = {2010}}

@article{Ji:2013dva,
	archiveprefix = {arXiv},
	author = {Ji, Xiangdong},
	doi = {10.1103/PhysRevLett.110.262002},
	eprint = {1305.1539},
	journal = {Phys. Rev. Lett.},
	pages = {262002},
	primaryclass = {hep-ph},
	slaccitation = {%%CITATION = ARXIV:1305.1539;%%},
	title = {{Parton Physics on a Euclidean Lattice}},
	volume = {110},
	year = {2013}}

@article{Zhang:2025hyo,
    author = "Zhang, Rui and Grebe, Anthony V. and Hackett, Daniel C. and Wagman, Michael L. and Zhao, Yong",
    title = "{Kinematically enhanced interpolating operators for boosted hadrons}",
    eprint = "2501.00729",
    archivePrefix = "arXiv",
    primaryClass = "hep-lat",
    reportNumber = "FERMILAB-PUB-24-0968-T",
    doi = "10.1103/6dh4-6k4t",
    journal = "Phys. Rev. D",
    volume = "112",
    number = "5",
    pages = "L051502",
    year = "2025"
}

@article{Ji:2020ect,
	archiveprefix = {arXiv},
	author = {Ji, Xiangdong and Liu, Yu-Sheng and Liu, Yizhuang and Zhang, Jian-Hui and Zhao, Yong},
	doi = {10.1103/RevModPhys.93.035005},
	eprint = {2004.03543},
	journal = {Rev. Mod. Phys.},
	number = {3},
	pages = {035005},
	primaryclass = {hep-ph},
	title = {{Large-momentum effective theory}},
	volume = {93},
	year = {2021}}

@article{Larin:1993tq,
      author         = "Larin, S. A.",
      title          = "{The Renormalization of the axial anomaly in dimensional
                        regularization}",
      journal        = "Phys. Lett.",
      volume         = "B303",
      year           = "1993",
      pages          = "113-118",
      doi            = "10.1016/0370-2693(93)90053-K",
      eprint         = "hep-ph/9302240",
      archivePrefix  = "arXiv",
      primaryClass   = "hep-ph",
      reportNumber   = "NIKHEF-H-92-18",
      SLACcitation   = "%%CITATION = HEP-PH/9302240;%%"
}

@article{Chen:2023lus,
    author = "Chen, Long",
    title = "{An observation on Feynman diagrams with axial anomalous subgraphs in dimensional regularization with an anticommuting \ensuremath{\gamma}$_{5}$}",
    eprint = "2304.13814",
    archivePrefix = "arXiv",
    primaryClass = "hep-ph",
    doi = "10.1007/JHEP11(2023)030",
    journal = "JHEP",
    volume = "2023",
    number = "11",
    pages = "30",
    year = "2023"
}

@article{Chen:2024hlv,
    author = "Chen, Long",
    title = "{A procedure g5anchor to anchor {\ensuremath{\gamma}}$_{5}$ in Feynman diagrams for the Standard Model}",
    eprint = "2409.08099",
    archivePrefix = "arXiv",
    primaryClass = "hep-ph",
    doi = "10.1007/JHEP05(2025)109",
    journal = "JHEP",
    volume = "05",
    pages = "109",
    year = "2025"
}

@article{Ahmed:2021spj,
    author = "Ahmed, Taushif and Chen, Long and Czakon, Micha\l{}",
    title = "{Renormalization of the flavor-singlet axial-vector current and its anomaly in dimensional regularization}",
    eprint = "2101.09479",
    archivePrefix = "arXiv",
    primaryClass = "hep-ph",
    reportNumber = "TTK-21-04, P3H-21-004",
    doi = "10.1007/JHEP05(2021)087",
    journal = "JHEP",
    volume = "05",
    pages = "087",
    year = "2021"
}

@article{Chen:2021gxv,
    author = "Chen, Long and Czakon, Micha\l{}",
    title = "{Renormalization of the axial current operator in dimensional regularization at four-loop in QCD}",
    eprint = "2112.03795",
    archivePrefix = "arXiv",
    primaryClass = "hep-ph",
    reportNumber = "TTK-21-54, P3H-21-099",
    doi = "10.1007/JHEP01(2022)187",
    journal = "JHEP",
    volume = "01",
    pages = "187",
    year = "2022"
}

@article{Chen:2022lun,
    author = "Chen, Long and Czakon, Micha{\l}",
    title = "{The MS renormalization constant of the singlet axial current operator at O({\ensuremath{\alpha}}s5) in QCD}",
    eprint = "2201.01797",
    archivePrefix = "arXiv",
    primaryClass = "hep-ph",
    reportNumber = "TTK-22-01, P3H-22-001",
    doi = "10.1016/j.physletb.2022.137266",
    journal = "Phys. Lett. B",
    volume = "832",
    pages = "137266",
    year = "2022"
}

@article{tHooft:1972tcz,
      author         = "'t Hooft, Gerard and Veltman, M. J. G.",
      title          = "{Regularization and Renormalization of Gauge Fields}",
      journal        = "Nucl. Phys.",
      volume         = "B44",
      year           = "1972",
      pages          = "189-213",
      doi            = "10.1016/0550-3213(72)90279-9",
      SLACcitation   = "%%CITATION = NUPHA,B44,189;%%"
}

@article{Breitenlohner:1975hg,
    author = "Breitenlohner, P. and Maison, D.",
    title = "{Dimensionally Renormalized Green's Functions for Theories with Massless Particles. 1.}",
    reportNumber = "MPI-PAE/PTh 15/75",
    doi = "10.1007/BF01609070",
    journal = "Commun. Math. Phys.",
    volume = "52",
    pages = "39",
    year = "1977"
}

@article{Breitenlohner:1976te,
    author = "Breitenlohner, P. and Maison, D.",
    title = "{Dimensionally Renormalized Green's Functions for Theories with Massless Particles. 2.}",
    reportNumber = "ITP-SB-76-19, NYU-TR2-76",
    doi = "10.1007/BF01609071",
    journal = "Commun. Math. Phys.",
    volume = "52",
    pages = "55",
    year = "1977"
}

@article{Breitenlohner:1977hr,
      author         = "Breitenlohner, P. and Maison, D.",
      title          = "{Dimensional Renormalization and the Action Principle}",
      journal        = "Commun. Math. Phys.",
      volume         = "52",
      year           = "1977",
      pages          = "11-38",
      doi            = "10.1007/BF01609069",
      SLACcitation   = "%%CITATION = CMPHA,52,11;%%"
}

@article{Breitenlohner:1983pi,
      author         = "Breitenlohner, Peter and Maison, Dieter and Stelle, K.
                        S.",
      title          = "{Anomalous Dimensions and the Adler-bardeen Theorem in
                        Supersymmetric {Yang-Mills} Theories}",
      journal        = "Phys. Lett.",
      volume         = "134B",
      year           = "1984",
      pages          = "63-66",
      doi            = "10.1016/0370-2693(84)90985-7",
      reportNumber   = "MPI-PAE/PTh 49/83",
      SLACcitation   = "%%CITATION = PHLTA,134B,63;%%"
}

@article{Luscher:2021bog,
    author = {L{\"u}scher, Martin and Weisz, Peter},
    title = "{Renormalization of the topological charge density in QCD with dimensional regularization}",
    eprint = "2103.15440",
    archivePrefix = "arXiv",
    primaryClass = "hep-ph",
    reportNumber = "CERN-TH-2021-041, MPP-2021-38",
    doi = "10.1140/epjc/s10052-021-09296-1",
    journal = "Eur. Phys. J. C",
    volume = "81",
    number = "6",
    pages = "519",
    year = "2021"
}

@article{Brodsky:1982gc,
    author = "Brodsky, Stanley J. and Lepage, G. Peter and Mackenzie, Paul B.",
    title = "{On the Elimination of Scale Ambiguities in Perturbative Quantum Chromodynamics}",
    reportNumber = "SLAC-PUB-3011, FERMILAB-PUB-83-040-T",
    doi = "10.1103/PhysRevD.28.228",
    journal = "Phys. Rev. D",
    volume = "28",
    pages = "228",
    year = "1983"
}

@article{Yang:2018nqn,
    author = "Yang, Yi-Bo and Liang, Jian and Bi, Yu-Jiang and Chen, Ying and Draper, Terrence and Liu, Keh-Fei and Liu, Zhaofeng",
    title = "{Proton Mass Decomposition from the QCD Energy Momentum Tensor}",
    eprint = "1808.08677",
    archivePrefix = "arXiv",
    primaryClass = "hep-lat",
    doi = "10.1103/PhysRevLett.121.212001",
    journal = "Phys. Rev. Lett.",
    volume = "121",
    number = "21",
    pages = "212001",
    year = "2018"
}

@article{CLQCD:2024yyn,
    author = "Du, Hai-Yang and others",
    collaboration = "CLQCD",
    title = "{Charmed meson masses and decay constants in the continuum limit from the tadpole improved clover ensembles}",
    eprint = "2408.03548",
    archivePrefix = "arXiv",
    primaryClass = "hep-lat",
    doi = "10.1103/PhysRevD.111.054504",
    journal = "Phys. Rev. D",
    volume = "111",
    number = "5",
    pages = "054504",
    year = "2025"
}

@article{supplemental,
    journal = "Supplemental Material"
}
\clearpage
\appendix

\begin{widetext}

\section{Supplemental Material}\label{sec:supplement}

\subsection{A. Perturbative results for matching coefficients $R_{ij}$ up to three-loops}\label{sec:Rexpressions}

According to Ref.~\cite{Pang:2024sdl}, converting the singlet axial-vector current $J_5^{\mu}$ and the topological current $K^{\mu}$ from the lattice $\mathrm{RI/MOM}$ scheme to the continuum $\overline{\mathrm{MS}}$ scheme requires a finite renormalization (matching) coefficient $R_{ij}$. This section details the perturbative calculation of $R_{ij}$ in dimensional regularization. Extending the results of Ref.~\cite{Pang:2024sdl}, we provide explicit expressions for $R_{ij}$ up to three-loop order. Note that our result for the one-loop perturbative matching coefficient differs from that reported in Ref.~\cite{Pang:2024sdl}, with the differences implied in the following discussion. Furthermore, we demonstrate that the off-diagonal elements of the (matching) coefficient can be neglected, leading to a simplified matching relation.

Following Ref.~\cite{Pang:2024sdl}, the following conditions apply within the RI/MOM renormalization scheme, namely
\begin{eqnarray}
    \label{eq:renorm_ori_sm}
    \begin{pmatrix} 
    K^{\mathrm{tree},g/q} \\ 
    J_5^{\mathrm{tree},g/q} 
    \end{pmatrix}
    =
    \begin{pmatrix} 
    Z_{11}^{\mathrm{RI}} & Z_{12}^{\mathrm{RI}} \\ Z_{21}^{\mathrm{RI}} & Z_{22}^{\mathrm{RI}} 
    \end{pmatrix}
    \begin{pmatrix} 
    K^{\mathrm{lat.},g/q} \\ 
    J_5^{\mathrm{lat.},g/q} 
    \end{pmatrix},
\end{eqnarray}
where we introduce the abbreviations $\mathcal{O}^{\mathrm{tree/lat.},{g/q}}\equiv\langle {g/q}|\mathcal{O}|{g/q}\rangle^{\mathrm{tree/lat.}}$ to simplify the notation. $|g\rangle$ or $|q\rangle$ denotes either gluon or quark external states with specific momentum and polarization. Note that $K^{\mathrm{tree},q}=J_5^{\mathrm{tree},g}=0$. Moreover, since only disconnected quark loops are involved, we have $J_5^{\mathrm{lat.},g}\approx0$. 
Thus, the RI/MOM renormalization constants can be determined as:
\begin{eqnarray}
    \label{eq:renorm_Z_sm}
    Z_{11}^{\mathrm{RI}}(\mu_{\rm RI}^2)&=&
    \frac{K^\mathrm{tree,g}}{K^{\mathrm{lat.,g}}},\quad\quad
    Z_{12}^{\mathrm{RI}}(\mu_{\rm RI}^2)=-
    \frac{K^{\mathrm{lat.,q}}}
    {J_5^{\mathrm{lat.,q}}}Z_{11}^{\mathrm{RI}},\nonumber\\
    Z_{21}^{\mathrm{RI}}(\mu_{\rm RI}^2)&=&0,\quad\quad\quad\quad\quad
    Z_{22}^{\mathrm{RI}}(\mu_{\rm RI}^2)=
    \frac{J_5^\mathrm{tree,q}}{J_5^{\mathrm{lat.,q}}},
\end{eqnarray}
where $\mu_{\rm RI}$ denotes the renormalization scale in the RI/MOM scheme. With the above approximation, $Z_{22}^{\mathrm{RI}}$ reduces to the axial-vector current renormalization constant $Z_A^\mathrm{RI}$ in Ref.~\cite{Hu:2023jet}.

In the spirit of $\mathrm{RI/MOM}$, the finite matching coefficients $R_{ij}$ can be defined in terms of the operator renormalization constants as follows:
\begin{equation} \label{eq:RmatrixDefinition}
\begin{pmatrix}
R_{11} &  R_{12} \\
R_{21} &  R_{22}
\end{pmatrix} 
\equiv 
\begin{pmatrix}
\bar{Z}^{\mathrm{DR}}_{11} &  \bar{Z}^{\mathrm{DR}}_{12} \\
\bar{Z}^{\mathrm{DR}}_{21} &  \bar{Z}^{\mathrm{DR}}_{22}
\end{pmatrix} 
\cdot 
\begin{pmatrix}
Z^{\mathrm{RI,\,DR}}_{11} &  Z^{\mathrm{RI,\,DR}}_{12} \\
Z^{\mathrm{RI,\,DR}}_{21} &  Z^{\mathrm{RI,\,DR}}_{22}
\end{pmatrix}^{-1} \Big|_{\epsilon \rightarrow 0} 
=  
\begin{pmatrix}
\bar{Z}^{\mathrm{lat}.}_{11} &  \bar{Z}^{\mathrm{lat}.}_{12} \\
\bar{Z}^{\mathrm{lat}.}_{21} &  \bar{Z}^{\mathrm{lat}.}_{22}
\end{pmatrix} 
\cdot 
\begin{pmatrix}
Z^{\mathrm{RI,\,lat}}_{11} &  Z^{\mathrm{RI,\,lat}}_{12} \\
Z^{\mathrm{RI,\,lat}}_{21} &  Z^{\mathrm{RI,\,lat}}_{22}
\end{pmatrix}^{-1} \,.  
\end{equation}
Consequently, the $R_{ij}$ determined using the first equality $\bar{Z}^{\mathrm{DR}} \cdot \big( Z^{\mathrm{RI,\,DR}} \big)^{-1} \big|_{\epsilon \rightarrow 0}$ with DR can be employed, together with the input $Z^{\mathrm{RI,\,lat}}_{ij} \equiv Z^{\mathrm{RI}}_{ij}$ computed in Eq.~\eqref{eq:renorm_Z_sm}, to obtain the values for the $\MSbar$-renormalization constants $\bar{Z}^{\mathrm{lat}.}_{ij}$ of the operator $K^{\mu}$ and $J_5^{\mu}$ but with UV-divergences regularized by the lattice, which can then be applied directly to the bare operator matrix elements computed on the lattice to obtain the $\MSbar$-renormalized results.
By virtue of Eq.~\eqref{eq:renorm_ori_sm}, we can further derive the following shortcut expressions for $R_{ij}$:  
\begin{equation} \label{eq:RmatrixShortcut}
\begin{pmatrix}
R_{11} &  R_{12} \\
R_{21} &  R_{22}
\end{pmatrix} 
= 
\begin{pmatrix}
\frac{ \overline{K}_{gg} }{K^{\mathrm{RI}}_{gg}} &  \frac{ \overline{K}_{\bar{q}q} }{J^{\mathrm{RI}}_{\bar{q}q}} \\
\frac{ \overline{J}_{gg} }{K^{\mathrm{RI}}_{gg}} &  \frac{ \overline{J}_{\bar{q}q} }{J^{\mathrm{RI}}_{\bar{q}q}}
\end{pmatrix} 
\end{equation}
where $\overline{K}(\overline{J})_{gg(\bar{q}q)}$ denote the Lorentz-invariant light-cone form-factors or matrix elements of the $\MSbar$-renormalized $K(J)$-current operator but with $\mathrm{RI/MOM}$-renormalized external $g(q)$-fields. Note that $K^{\mathrm{RI}}_{gg}$ and $J^{\mathrm{RI}}_{\bar{q}q}$ are equal to, respectively, the tree-level values of the aforementioned matrix elements of $K(J)$-current operator as a result of the $\mathrm{RI/MOM}$ renormalization condition Eq.~\eqref{eq:renorm_ori_sm}.

Extra care is required in the calculation of the dimensionally-regularized bare matrix elements and the subsequent extraction of renormalization constants entering the r.h.s. of Eq.\eqref{eq:RmatrixDefinition} and Eq.\eqref{eq:RmatrixShortcut}, because of the notorious issue in treating $\gamma_5$, and to a lesser extent $\epsilon^{\mu\nu\rho\sigma}$ in DR, especially in the present case due to the involvement of the axial anomaly. 
In view of the applications related to the study of polarized structure functions of the proton, we choose a (modified) $\MSbar$ prescription where the gauge invariance and Adler-Bell-Jackiw equation are preserved (hence without chiral symmetry). 
So far, there is no practical prescription known to the authors to apply consistently an anticommuting $\gamma_5$ in DR for computing loop amplitudes with (closed) fermion chains with an odd number of $\gamma_5$, especially when the amplitudes are anomalous~\cite{Chen:2023lus,Chen:2024hlv}.
On the other hand, the so-called Larin’s prescription~\cite{Larin:1993tq} of $\gamma_5$, where $\gamma_5$ in DR is treated non-anticommuting~\cite{tHooft:1972tcz,Breitenlohner:1977hr,Breitenlohner:1975hg,Breitenlohner:1976te} while the spacetime-metric tensors resulting from contracting a pair of $\epsilon^{\mu\nu\rho\sigma}$ are
all in D-dimensions, is particularly suited for dealing with the QCD corrections at hand: 
both gauge invariance and the Adler-Bell-Jackiw equation 
are maintained and it is technically convenient to apply at high loop orders in QCD.

With this set-up, we have derived the perturbative expressions for the bare matrix elements involved in Eq.\eqref{eq:RmatrixShortcut} up to three loops, truncated to high orders in $\epsilon$ sufficient for obtaining their renormalized counterparts to $\mathcal{O}(\epsilon^0)$; 
they are subsequently employed to extract the perturbative results for finite $R_{ij}$ according to both Eq.\eqref{eq:RmatrixShortcut} and Eq.\eqref{eq:RmatrixDefinition}, with perfect agreement found between the two. 
(For more technical details of the computational workflow and evaluation of loop integrals, we kindly refer to Refs.~\cite{Ahmed:2021spj,Chen:2021gxv}.) 
For the sake of references, we document below the perturbative results for the finite matching coefficients $R_{ij}$ up to three loops in Landau gauge.

We begin with $R_{11}$ that was utilized in the present work, which reads 
\begin{eqnarray} \label{eq:R11exp}
R_{11} &=& 1 \,+\,     
a_s \, 
\Big( C_A \, \left(\frac{11 L_{\mu }}{3}+\frac{367}{36}\right) 
\,+\, 
n_f \, \left(-\frac{2 L_{\mu }}{3}-\frac{10}{9}\right) 
\Big) \nonumber\\
&+&
a_s^2 \, 
\Big(
C_A^2 \, \left(\frac{15 \zeta _3}{4}+\frac{121 L_{\mu }^2}{9}+\frac{4649 L_{\mu }}{54}+\frac{99047}{648}\right)
\,+\, 
C_A n_f \, \left(-16 \zeta _3-\frac{44 L_{\mu }^2}{9}-\frac{677 L_{\mu }}{27}-\frac{2762}{81}\right)
\nonumber\\ &+& 
C_F n_f \, \left(16 \zeta _3-8 L_{\mu }-\frac{161}{6}\right)
\,+\, 
n_f^2 \, \left(\frac{4 L_{\mu }^2}{9}+\frac{40 L_{\mu }}{27}+\frac{100}{81}\right)
\Big) \nonumber\\
&+&
a_s^3 \, 
\Big(
C_A^3 \, \left(-\frac{3743 \zeta _3}{288}-\frac{31475 \zeta _5}{192}+\frac{165 \zeta _3 L_{\mu }}{4}+\frac{1331 L_{\mu }^3}{27}+\frac{55627 L_{\mu }^2}{108}+\frac{1273537 L_{\mu }}{648}+\frac{17627903}{5832}\right)
\nonumber\\ &+& 
C_A^2 n_f \, \left(-\frac{16627 \zeta _3}{36}+\frac{470 \zeta _5}{3}-\frac{367 \zeta _3 L_{\mu }}{2}-\frac{242 L_{\mu }^3}{9}-\frac{2194 L_{\mu }^2}{9}-\frac{28805 L_{\mu }}{36}-\frac{2061755}{1944}\right)
\nonumber\\ &+& 
C_A C_F n_f \, \left(\frac{1306 \zeta _3}{3}+120 \zeta _5+176 \zeta _3 L_{\mu }-\frac{187 L_{\mu }^2}{3}-\frac{2735 L_{\mu }}{6}-\frac{331903}{324}\right)
\nonumber\\ &+& 
C_A n_f^2 \, \left(\frac{524 \zeta _3}{9}+32 \zeta _3 L_{\mu }+\frac{44 L_{\mu }^3}{9}+\frac{319 L_{\mu }^2}{9}+\frac{4895 L_{\mu }}{54}+\frac{89833}{972}\right)
\nonumber\\ &+& 
C_F^2 n_f \, \left(\frac{460 \zeta _3}{3}-240 \zeta _5+19 L_{\mu }+\frac{1333}{18}\right)
\,+\,
C_F n_f^2 \, \left(-72 \zeta _3-32 \zeta _3 L_{\mu }+\frac{34 L_{\mu }^2}{3}+\frac{202 L_{\mu }}{3}+\frac{20375}{162}\right)
\nonumber\\ &+& 
n_f^3 \, \left(-\frac{8 L_{\mu }^3}{27}-\frac{40 L_{\mu }^2}{27}-\frac{200 L_{\mu }}{81}-\frac{1000}{729}\right)
\Big) \,+\, \mathcal{O}(a_s^4)\,,
\end{eqnarray}
where the shorthand notations $a_s \equiv \frac{\alpha_s}{4\, \pi}$ and $L_{\mu } \equiv \ln \big(\frac{{\mu}^2}{\mu_{\rm RI}^2} \big)$ are introduced.
The definitions of the quadratic Casimir color constants involved above are as usual: $C_A = N_c \,, \, C_F = (N_c^2 - 1)/(2 N_c) \,$ with $N_c =3$ in QCD and we have set the color-trace normalization factor to its value ${1}/{2}$.
($\zeta_n$ is the usual Riemann $\zeta(n)$ function.)
The perturbative corrections to $R_{22}$ starts only from two-loops and are simpler:  
\begin{eqnarray} \label{eq:R22exp}
R_{22} &=& 1 \,+\, 
a_s^2 \, 
\Big(
C_F n_f \, \left(-6 L_{\mu }-12\right)
\Big) \nonumber\\
&+&
a_s^3 \, 
\Big(
C_A C_F n_f \, \left(66 \zeta _3-22 L_{\mu }^2-\frac{406 L_{\mu }}{3}-\frac{4643}{18}\right)
\nonumber\\ &+& 
C_F^2 n_f \, \left(-48 \zeta _3+18 L_{\mu }+\frac{77}{2}\right) 
\,+\, 
C_F n_f^2 \, \left(4 L_{\mu }^2+\frac{52 L_{\mu }}{3}+\frac{289}{9}\right) 
\Big) 
\,+\, \mathcal{O}(a_s^4)\,.
\end{eqnarray}
The results for the off-diagonal elements are:
\begin{eqnarray} \label{eq:R12exp}
R_{12} &=& 1 \,+\, 
a_s \, 
\Big(
C_F \, \left(3 L_{\mu }+6\right)
\Big) \nonumber\\
&+&
a_s^2 \, 
\Big(
C_A C_F \, \left(-33\zeta _3+11 L_{\mu }^2+\frac{203 L_{\mu }}{3}+\frac{4643}{36}\right)
\,+\, C_F^2 \, \left(24 \zeta _3-9 L_{\mu }-\frac{77}{4}\right)
\,+\, 
C_F n_f \, \left(-2 L_{\mu }^2-\frac{26 L_{\mu }}{3}-\frac{289}{18}\right)
\Big) \nonumber\\
&+&
a_s^3 \, 
\Big(
C_A^2 C_F \, \left(-\frac{14419 \zeta _3}{12}+\frac{1045 \zeta _5}{4}-363 \zeta _3 L_{\mu }+\frac{121 L_{\mu }^3}{3}+\frac{2437 L_{\mu }^2}{6}+\frac{30395 L_{\mu }}{18}+\frac{1691641}{648}\right)
\nonumber\\ &+& 
C_A C_F^2 \, \left(668 \zeta _3+264 \zeta _3 L_{\mu }-\frac{99 L_{\mu }^2}{2}-327 L_{\mu }-\frac{104701}{216}\right)
\nonumber\\ &+& 
C_A C_F n_f \, \left(\frac{140 \zeta _3}{3}-6 \zeta _3 L_{\mu }-\frac{44 L_{\mu }^3}{3}-\frac{376 L_{\mu }^2}{3}-\frac{4148 L_{\mu }}{9}-\frac{2 \pi ^4}{5}-\frac{102293}{162}\right)
\nonumber\\ &+& 
C_F^3 \, \left(64 \zeta _3-240 \zeta _5+\frac{63 L_{\mu }}{2}+\frac{707}{12}\right)
\,+\,
C_F^2 n_f \, \left(72 \zeta _3+24 \zeta _3 L_{\mu }-6 L_{\mu }^2-75 L_{\mu }+\frac{2 \pi ^4}{5}-\frac{12899}{54}\right)
\nonumber\\ &+& 
C_F n_f^2 \, \left(\frac{8 \zeta _3}{3}+\frac{4 L_{\mu }^3}{3}+\frac{26 L_{\mu }^2}{3}+\frac{250 L_{\mu }}{9}+\frac{5005}{162}\right)
\Big) 
\,+\, \mathcal{O}(a_s^4)\,,
\end{eqnarray} 
and 
\begin{eqnarray} \label{eq:R21exp}
R_{21} &=& 1 \,+\, 
a_s \, 
n_f \, \left(-2\right)
\,+\,
a_s^2 \, 
\Big(
C_A n_f \, \left(-\frac{22 L_{\mu }}{3}-\frac{367}{18}\right)
\,+\,
n_f^2 \, \left(\frac{4 L_{\mu }}{3}+\frac{20}{9}\right)
\Big) \nonumber\\
&+&
a_s^3 \, 
\Big(
C_A^2 n_f \, \left(-\frac{15 \zeta _3}{2}-\frac{242 L_{\mu }^2}{9}-\frac{4649 L_{\mu }}{27}-\frac{99047}{324}\right)
\,+\,
C_A n_f^2 \, \left(32 \zeta _3+\frac{88 L_{\mu }^2}{9}+\frac{1354 L_{\mu }}{27}+\frac{5524}{81}\right)
\nonumber\\ &+& 
C_F n_f^2 \, \left(-32 \zeta _3+16 L_{\mu }+\frac{161}{3}\right)
\,+\,
n_f^3 \, \left(-\frac{8 L_{\mu }^2}{9}-\frac{80 L_{\mu }}{27}-\frac{200}{81}\right)
\Big) 
\,+\, \mathcal{O}(a_s^4)\,.
\end{eqnarray} 

The dependence of $R_{ij}$ on ${\mu}^2$ exhibited in the above perturbative expressions are governed by their RG equations.
Following from the definition Eq.~\eqref{eq:RmatrixDefinition} the RG equations for the matching coefficients $R_{ij}$ read  
\begin{eqnarray}
\label{eq:AMDmatrix}
{\mu}^2\, \frac{\mathrm{d}}{\mathrm{d}\, {\mu}^2 }\,
\begin{pmatrix}
R_{11} &  R_{12} \\
R_{21} &  R_{22}
\end{pmatrix} 
= 
\begin{pmatrix}
\gamma_{11} &  \gamma_{12} \\
0 &  \gamma_{22}
\end{pmatrix}
\cdot 
\begin{pmatrix}
R_{11} &  R_{12} \\
R_{21} &  R_{22}
\end{pmatrix} ,
\end{eqnarray}
where the matrix of anomalous dimensions on the r.h.s.~is fully determined by the (proper or modified) $\MSbar$ renormalization constants in Eq.~\eqref{eq:RmatrixDefinition} in Larin’s prescription~\cite{Larin:1993tq}; 
these anomalous dimensions observe the relations~\cite{Breitenlohner:1983pi,Ahmed:2021spj,Luscher:2021bog} $\gamma_{11} = -\frac{\mathrm{d} \ln a_s}{\mathrm{d} \ln {\mu}^2} = -\beta $ and $\gamma_{22} = -2 a_s\,  n_f \, \gamma_{12} $  known now to four~\cite{Chen:2021gxv} and five loops~\cite{Chen:2022lun}.
~\\

With the explicit expression Eq.~\eqref{eq:R11exp} for $R_{11}$, it is interesting to observe that in the limit of large $\beta_0$ or $n_f$, 
the leading terms in powers of $\mathcal{O}(a^n_s\, \beta^n_0)$ exhibit the following neat pattern: 
\begin{equation}\label{eq:LNFlimit}
R_{11}\Big|_{\text{large $\beta_0$}} 
\approx 
1 \,+\, a_s\, \beta_0 \, \big( L_{\mu} + \frac{5}{3} \big)     
\,+\, a_s^2\, \beta_0^2\, \big( L_{\mu} + \frac{5}{3} \big)^2
\,+\, a_s^3\, \beta_0^3 \,\big( L_{\mu} + \frac{5}{3} \big)^3
\,+\, \mathcal{O}(a_s^4)\,,
\end{equation}
which seems to comply with the geometric series $1 \,+\, \sum_{1}^{\infty} a_s^n\, \beta_0^n \big( L_{\mu} + \frac{5}{3} \big)^n$.
Consequently, taking the scale $L_{\mu} = \ln \big(\frac{{\mu}^2}{\mu_{\rm RI}^2} \big) = -\frac{5}{3}$, namely $\mu = e^{-\frac{5}{6}}\, \mu_{\rm RI}$, can eliminate completely all these pieces from the fixed-order expression for $R_{11}$, and hence could improve the convergence of the first few terms of the truncated perturbative series (at least for $\mu > 2\,\mathrm{GeV}$). 
This expectation is supported by examining the $\mu$-variation of the ratios of the pure three- and two-loop corrections to $R_{11}$ to its results at the previous orders (with $\mu_{\rm RI}^2$ around, e.g.,~$10\ \mathrm{GeV}^2$ and larger).
Interestingly, this choice of the optimal scale happens to coincide with the original or standard Brodsky-Lepage-Mackenzie  scale-setting~\cite{Brodsky:1982gc}.
Alternatively, for the low scale region of $\mu_{\rm RI}$,~e.g.~below $2\,\mathrm{GeV}$, one may simply choose to take the fixed-order expression of $R_{11}(\mu,\,\mu_{\rm RI})$ with $\alpha_s(\mu)$ evaluated directly at the relatively high reference scale $\mu^2 = 10\,\mathrm{GeV}^2$, irrespective of $\mu_{\rm RI}$. 
At the $\overline{\mathrm{MS}}$ scale $\mu^2 = 10,\mathrm{GeV}^2$, Fig.~\ref{fig:matching_running} (left) depicts the $\mu_{\mathrm{RI}}$ dependence of all four matching coefficients $R_{ij}$ at two and three loops, while Fig.~\ref{fig:matching_running} (right) focuses on the evolution of $R_{11}$ from one to three loops, obtained via performing RG running from the fixed-order results at the intermediate optimal scale $\mu_i = e^{-\frac{5}{6}}\, \mu_{\rm RI}$.
These plots show that the $R_{11}(\mu,\,\mu_{\rm RI})$ evaluated in this way exhibit acceptable convergence behaviors in the region of $a^2\,p^2$ used. 
Furthermore, the off-diagonal matching coefficient $R_{12}$ is much less than $R_{11}$ across the region of $a^2\,p^2$ in question, almost an order of magnitude less for $a^2p^2 > 4$, which --- together with the information on the bare matrix elements discussed previously --- supports its omission from the present analysis.

\begin{figure*}[pt] 
   \centering
   \begin{tabular}{cc}
       \includegraphics[width=0.47\textwidth]{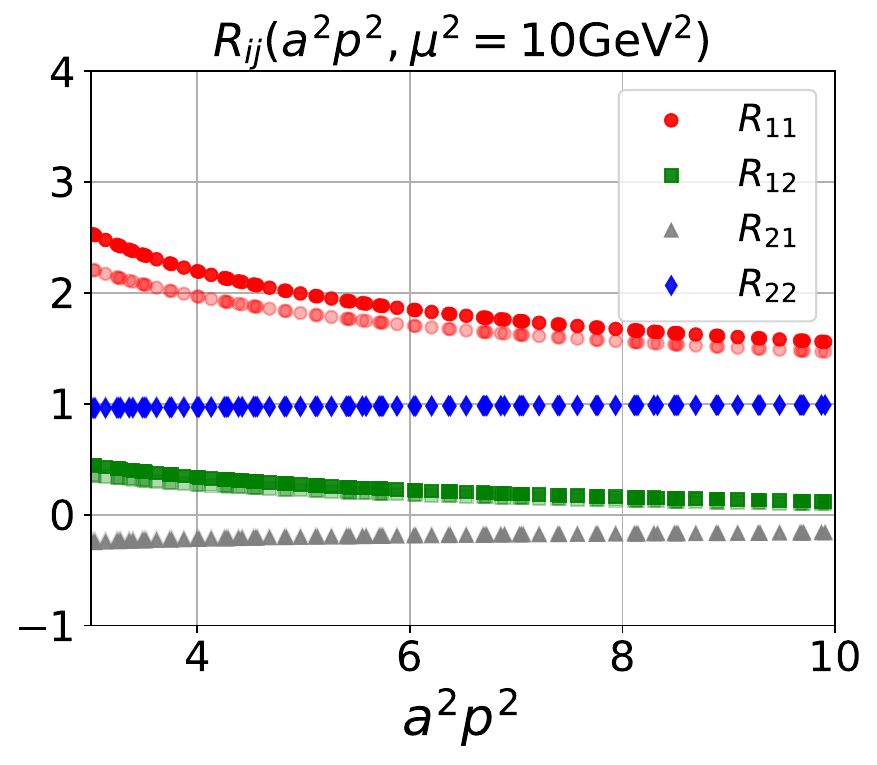}& 
       \includegraphics[width=0.47\textwidth]{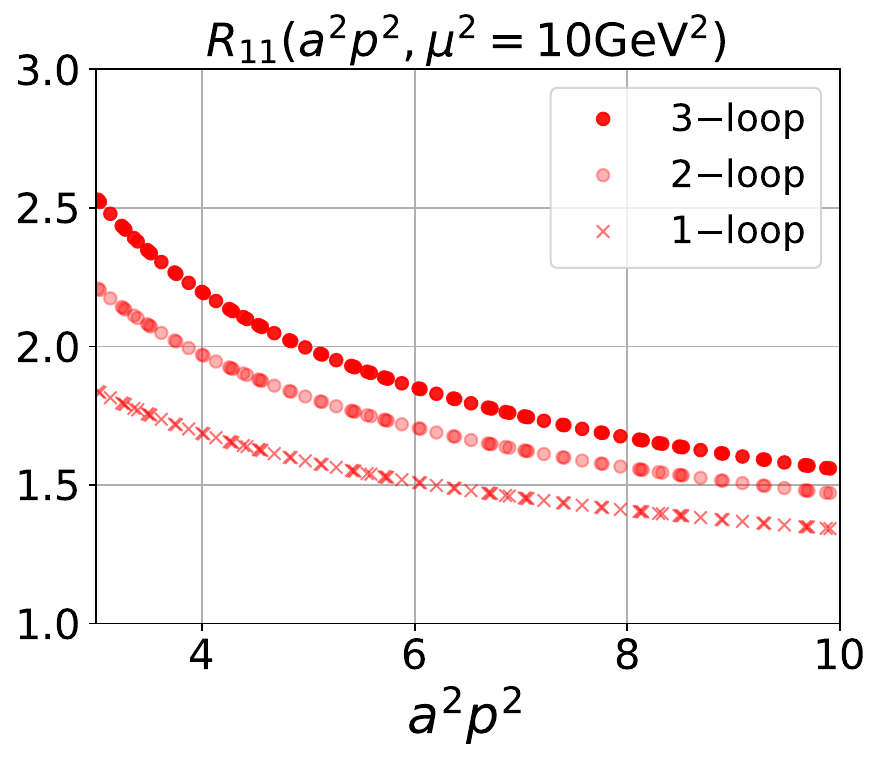}\\
   \end{tabular}
   \caption{Matching coefficients in the $\mathrm{RI/MOM}$ scale at ${\mu}^2 = 10\ \mathrm{GeV}^2$. The matching coefficients $R_{ij}$ (two- and three-loop, in light and dark color, respectively) are shown in the left panel; the right panel specifically presents $R_{11}$ from one to three loops. In order to combine with lattice data, we change the horizontal axis to the dimensionless $\mathrm{RI/MOM}$ scale $a^2p^2$. }
    \label{fig:matching_running}
\end{figure*}

Furthermore, motivated by the features observed in Fig.~\ref{fig:matching_running}, it can be demonstrated that the off-diagonal elements of the matching coefficient are negligible. This leads to a significant simplification of the matching relations. Starting from simplified RI/MOM-renormalized gluon helicities $\Delta G$ and its $\overline{\mathrm{MS}}$ matching, namely
\begin{eqnarray}
    \label{eq:deltaGmsri_sm}
    \Delta G^{\mathrm{RI}}&=&Z_{11}^{\mathrm{RI}}
    \big(\Delta G^\mathrm{B.}-
    \frac{K^{\mathrm{lat.,q}}}{J_5^{\mathrm{lat.,q}}
    }\Delta\Sigma^{\mathrm{B.}}\big),\nonumber\\
    \Delta G^{\overline{\mathrm{MS}}}&=&R_{11}\Delta G^{\mathrm{RI}}+R_{12}\Delta\Sigma^{\mathrm{RI}},
\end{eqnarray}
we can obtain the exact formulations for gluon helicity in the $\overline{\mathrm{MS}}$ scheme
\begin{eqnarray}
    \Delta G^{\overline{\mathrm{MS}}}(a^2p^2,{\mu}^2)&=&R_{11}(a^2p^2,{\mu}^2)Z_{11}^{\mathrm{RI}}(a^2p^2)\big(\Delta G^{\mathrm{B.}}-\frac{K^{\mathrm{lat.,q}}}{J_5^{\mathrm{lat.,q}}}(a^2p^2)\Delta\Sigma^{\mathrm{B.}}\big)\nonumber\\
    &&+R_{12}(a^2p^2,{\mu}^2)Z_A^{\mathrm{RI}}(a^2p^2)\Delta\Sigma^{\mathrm{B.}},
\end{eqnarray}
where $Z_{11}^{\mathrm{RI}}=
K^\mathrm{tree,g}/K^{\mathrm{lat.,g}}$ is the RI/MOM diagonal renormalization constant of gluon helicity, and we introduce the abbreviations $\mathcal{O}^{\mathrm{tree/lat.},{g/q}}\equiv\langle {g/q}|\mathcal{O}|{g/q}\rangle^{\mathrm{tree/lat.}}$ to simplify the notation. In addition, we express the dependence on the squared renormalization scale $\mu_{\rm RI}^2$ of the $\mathrm{RI/MOM}$ scheme in terms of the dimensionless quantity $a^2p^2$, and the same below. We now implement a three-loop approximation for the ratio term $K^{\mathrm{lat,q}}/{J_5^{\mathrm{lat,q}}}$. Noting that $R_{12}$ can be determined from the topological current $K^\mu$ for quark external states while $R_{22}\approx1$ is fixed by the axial vector for quark external states. Finally we approximate 
\begin{eqnarray}
    \big(\bar{Z}_{11}^{\mathrm{lat}.}R_{12}\big)(a^2p^2\rightarrow0,{\mu}^2)
    &\approx&\big(\bar{Z}_{11}^{\mathrm{lat}.}\frac{R_{12}}{R_{22}}\big)(a^2p^2\rightarrow0,{\mu}^2)\nonumber\\
    &=&\big(\bar{Z}_{11}^{\mathrm{lat}.}\frac{K^{3\mathrm{-loop,q}}}{J_5^{3\mathrm{-loop,q}}}\big)(a^2p^2\rightarrow0,{\mu}^2)\nonumber\\
    &\approx&
    \big(\bar{Z}_{11}^{\mathrm{lat}.}\frac{K^{\mathrm{lat,q}}}{J_5^{\mathrm{lat,q}}}\big)(a^2p^2\rightarrow0,{\mu}^2).
\end{eqnarray}
where we used a redefined expression $\bar{Z}_{11}^{\mathrm{lat}.}\equiv R_{11}Z_{11}^{\mathrm{RI}}$. 
Furthermore, owing to the protection by chiral symmetry, the matching coefficient for the axial-vector current remains unity up to two loops, i.e., $\bar{Z}_A^{\mathrm{lat}} \equiv Z_A^{\overline{\mathrm{MS}}} \approx Z_A^{\mathrm{RI}}$.
This convention will be consistently maintained throughout the subsequent discussion. This approximation enables the following simplified expression for gluon helicity at three-loop order:
\begin{eqnarray}
    \Delta G^{\overline{\mathrm{MS}}}&\overset{3\mathrm{-loop}}{\approx}&
    \bar{Z}_{11}^{\mathrm{lat}.}\big(\Delta G^{\mathrm{B.}}-R_{12}\Delta\Sigma^{\mathrm{B}.}\big)+R_{12}\bar{Z}_A^{\mathrm{lat}.}\Delta\Sigma^{\mathrm{B.}}\nonumber\\
    &=&\bar{Z}_{11}^{\mathrm{lat}.}\Delta G^{\mathrm{B.}}
    +(\bar{Z}_A^{\mathrm{lat}.}-
    \bar{Z}_{11}^{\mathrm{lat.}})
    R_{12}\Delta\Sigma^{\mathrm{B.}}\nonumber\\
    &\approx& \bar{Z}_{11}^{\mathrm{lat}.}\Delta G^{\mathrm{B.}},
\end{eqnarray}
where the final reduction is justified based on the hierarchy $\bar{Z}_{11}^{\mathrm{lat}.} \gg (\bar{Z}_A^{\mathrm{lat}.}-\bar{Z}_{11}^{\mathrm{lat}.}) R_{12}$, and by the fact that $\Delta G^{\mathrm{B.}}$ is comparable in magnitude to $\Delta \Sigma^{\mathrm{B.}}$. The above arguments are all self-evident in the lattie QCD calculations.
The derivation for the expression $\Delta\Sigma^{\overline{\mathrm{MS}}}\approx \bar{Z}_{A}^{\mathrm{lat}.}\Delta\Sigma^\mathrm{B.}$ parallels the arguments used in the approximation above.

\subsection{B. Bare nucleon matrix elements with topological current in the moving frame}\label{sec:BME_extract}

The results in this work, are based on the $n_f=2+1$ flavor ensembles from the CLQCD collaboration using the tadpole improved tree level Symanzik (TITLS) gauge action and the tadpole improved tree level Clover (TITLC) fermion action~\cite{Hu:2023jet}.

The TITLS gauge action is defined in the following,
\begin{align}
\label{eq:gauge_action}
    S_g = \frac{1}{N_c} \mathrm{Re} \sum_{x,\mu<\nu}\mathrm{Tr} 
    \left[ 
        1- \hat{\beta} 
        \left(
            \mathcal{P}^U_{\mu,\nu}(x)+\frac{c_1\mathcal{R}^U_{\mu,\nu}(x)}{1-8c_1^0}
        \right)
    \right],
\end{align}
and  
\begin{align}
    \mathcal{P}^{U}_{\mu,\nu}(x)& = U_\mu(x)U_\nu(x+a\hat{\mu})U^{\dagger}_\mu(x+a\hat{\nu})U^{\dagger}_\nu(x), \notag
    \\
    \mathcal{R}^{U}_{\mu,\nu}(x)& = U_\mu(x) U_\mu(x+a\hat{\mu}) U_\nu(x+2a\hat{\mu})U^{\dagger}_\mu(x+a\hat{\mu}+a\hat{\nu}) U^{\dagger}_\mu(x+a\hat{\nu}) U^{\dagger}_\nu(x), \notag
    \\
    U_{\mu}(x) &= P
    \left[ 
        \mathrm{exp} 
        \left( 
            {\rm i} g_0 \int_{x}^{x+\hat{\mu}a} \mathrm{d}y A_{\mu}(y)
        \right) 
    \right], \notag
\end{align}
$\hat{\beta}=(1-8c_1^0)\frac{6}{g_0^2u_0^4}\equiv 10/(g_0^2u_0^4)$ with $c_1^0=-\frac{1}{12}$, $c_1=\frac{c_1^0}{ u_0^2}$, $u_0=\langle \frac{\mathrm{Re}\mathrm{Tr}\sum_{x,\mu<\nu}\mathcal{P}^{U}_{\mu\nu}(x)}{6N_c\tilde{V}} \rangle^{1/4}$ is the tadpole improvement factor.

The TITLC fermion action uses 1-step stout smeared link $V$ with smearing parameter $\rho=0.125$,

\begin{align}
    \label{eq:quark_action}
    &S_q(m)=\sum_{x,\mu=1,...,4,\eta=\pm}\bar{\psi}(x)\sum\frac{1+\eta\gamma_{\mu}}{2}V_{\eta\mu}(x)\psi(x{+\eta\hat{\mu}a})\nonumber
    \\ 
    &\quad +\sum_x\psi(x)
    \left[
        -(4+ma) \delta_{y,x} + c_{\rm sw} \sigma^{\mu\nu} g_0 F_{\mu\nu}
    \right]
    \psi(x),
\end{align}
where $c_{\rm sw}=\frac{1}{v^3_{0}}$ with $v_0=\langle \frac{\mathrm{Re}\mathrm{Tr}\sum_{x,\mu<\nu}\mathcal{P}^{V}_{\mu\nu}(x)}{6N_c\tilde{V}} \rangle^{1/4}$, and the field strength tensor $F_{\mu\nu}(x)$ is defined through the following discretized formulations
\begin{eqnarray}
    \label{eq:fmunu}
    F_{\mu\nu}(x) &=& \frac{i}{8g_0a^2} (\mathcal{P}_{\mu,\nu}^U(x)-\mathcal{P}_{\nu,\mu}^U(x)+\mathcal{P}_{\nu,-\mu}^U(x)-\mathcal{P}_{-\mu,\nu}^U(x) \nonumber \\
    && + \mathcal{P}_{-\mu,-\nu}^U(x) -\mathcal{P}_{-\nu,-\mu}^U(x) + \mathcal{P}_{-\nu,\mu}^U(x) -\mathcal{P}_{\mu,-\nu}^U(x)).
\end{eqnarray}
The detailed information for the simulation is given in Table~\ref{tab:ensem_total}. For all ensembles except C48P23, the quark propagators are computed for all time slices using the distillation smearing method~\cite{HadronSpectrum:2009krc}, 1-step stout smearing is applied to the links in the Laplacian operator (with smearing parameter of $\rho=0.125$), and number of eigenvectors $N_{\mathrm{ev}}$ in the smearing operator is 100. For the proton external states with momentum greater than $0.5$ GeV, momentum smearing~\cite{Bali:2016lva,Egerer:2020hnc} is further implemented using smearing parameter $\textbf{k} = (0, 0, 2)$. The C48P23 ensemble is intended specifically for renormalization calculations; therefore, neither distillation nor momentum smearing is employed.

\begin{table}[hbt!]
\centering
\resizebox{0.60\columnwidth}{!}{
\begin{tabular}{|c|c|c|c|c|c|c|c|c|} 
\hline
Symbol & $\hat{\beta}$& $a$ (fm) & $\tilde{L}^3\times \tilde{T}$ & $m_{\pi}$ (MeV) & $N_{\mathrm{cfg}}$ & $\rho$ & $N_{ev}$ & $\textbf{k}$\\
\hline 
C24P29 & \multirow{2}{*}{6.200} & \multirow{2}{*}{0.10524(05)(62)} & $24^3\times 72$ & 292.3(1.0) & 780 & 0.125 & 100 & $(0,0,2)$\\
C48P23 & & & $48^3\times 96$ & 224.1(1.2) & 400 & 0.125 & - & -\\
\hline
E32P29 & 6.308 & 0.08973(20)(53) & $32^3\times 64$ & 287.3(2.5) & 890 & 0.125 & 100 & $(0,0,2)$\\
\hline
F32P30 & 6.410 & 0.07753(03)(45) & $32^3\times 96$ & 300.4(1.2) & 800 & 0.125 & 100 & $(0,0,2)$\\
\hline
\end{tabular}
}
\caption{Lattice setup used for the simulation, which including gauge coupling $\hat{\beta}=10/(g_0^2u_0^4)$, lattice spacing $a$, dimensionless lattice size $\tilde{L}^3\times \tilde{T}$, corresponding pion mass $m_{\pi}$, number of configurations $N_{\mathrm{cfg}}$, 1-step stout smearing parameter $\rho$, number of distillation eigenvectors $N_{\mathrm{ev}}$ and momentum smearing parameter $\textbf{k}$.} 
\label{tab:ensem_total}
\end{table}

In lattice computations of the topological current
\begin{equation}
    \label{eq:tp_current_sm}
    K^\mu(x)=\epsilon^{\mu\nu\rho\sigma}\text{Tr}[A_\nu F_{\rho\sigma}-\frac23ig_0A_\nu A_\rho A_\sigma](x),
\end{equation}
the fundamental components requiring evaluation are the gauge potential $A_\mu(x)$ and the field strength tensor $F_{\mu\nu}(x)$, where $A_\mu(x)$ is defined through the following discretized formulations
\begin{equation}
    \label{eq:amu}
    A_\mu(x+\hat\mu/2)=\left[\frac{U_\mu(x)-U_\mu^\dagger(x)}{2 i g_0 a}\right]_{\mathrm {traceless}.}.
\end{equation}
While an alternative full-lattice definition $\tilde{A}_\mu(x) = [A_\mu(x + \hat{\mu}/2)+A_\mu(x - \hat{\mu}/2)]/2$ exists, the half-lattice and full-lattice forms of the gauge potential differ only in their discretization errors. We therefore use the half-lattice definition in all following calculations. Based on definition of the local topological current $K^\mu$ in Eq.~\eqref{eq:tp_current_sm}, its temporal and spatial components read explicitly: 
\begin{eqnarray}
    \label{eq:tp_ti}
    K_i&\equiv&K^{E\times A}_i+K_i^{A_tB}+K_i^{3A}=2\text{Tr}[(E\times A)_i] - 2\text{Tr}[A_t B_i] + \frac{2}{3}ig_0\epsilon^{ijk}\mathrm{Tr}[A_t [A_j, A_k]]\nonumber\\
    K_t&\equiv&K_t^{B\cdot A}+K_t^{3A}=2\text{Tr}[B\cdot A]-\frac{2}{3}ig_0\epsilon^{ijk}\mathrm{Tr}[A_i A_j A_k],
\end{eqnarray}
where $E_i=F_{ti}$ and $B_i=F_{jk}$. Note that the component expression of $K^\mu$ is defined in Euclidean space. Due to signal-to-noise ratio considerations, 5-HYP smearing is applied to the topological current throughout this simulation.

Following Ref.~\cite{Pang:2024sdl}, the total gluon helicity $\Delta G$ is obtained from the matrix element of the local topological current,
\begin{equation}
    \label{eq:threept_sm}
    \frac{\langle \mathrm{PS}_{\mathrm{Proj.i}}|\int d^3x K^{\mu}_{\mathrm{C.G.}}(x)|\mathrm{PS}_{\mathrm{Proj.i}}\rangle}{2S^{\mu}\langle\mathrm{PS}_{\mathrm{Unproj.}}|\mathrm{PS}_{\mathrm{Unproj.}}\rangle}|_{\mathrm{IMF}}= \Delta G.
\end{equation}
Here, $| \mathrm{PS}_{\mathrm{Unproj.}} \rangle$ denotes a proton external state with definite momentum and no polarization, whereas $| \mathrm{PS}_{\mathrm{Proj.,i}} \rangle$ refers to the corresponding state polarized along the spatial direction $i$. The component $S^{\mu}$ is the $\mu$th element of the polarization vector associated with direction $i$. The subscript $\mathrm{C.G.}$ indicates that the topological current is fixed in the Coulomb gauge, while the subscript of IMF indicates that matrix elements need to be extrapolated to the infinite momentum frame (IMF). We employ the distillation smearing technique~\cite{HadronSpectrum:2009krc} for proton external states at low momenta ($p \leq 0.5 \ \mathrm{GeV}$). For states with higher momentum, we combine distillation with momentum smearing~\cite{Bali:2016lva,Egerer:2020hnc}.

By defining the enhanced proton operator~\cite{Zhang:2025hyo}
\begin{equation}
    N(\vec{p},t)=\sum_{\vec{x}}e^{-i\vec{p}\cdot\vec{x}}P_+(u^TC\gamma_5\gamma_tdu)(\vec{x},t),
\end{equation}
the two-point correlation function (2pt) can be defined as
\begin{equation}
    \label{eq:2pt_sm}
    C_2(p,t_f)=\sum_{t_s}\langle 0|N(p,t_s+t_f)N^\dagger(p,t_s)|0\rangle,
\end{equation}
with source time slice $t_s$, difference between source and sink time slice $t_f$. Extracting the effective mass from two-point correlation functions is an essential procedure, as it isolates the ground-state contribution—a prerequisite for reliably determining baryon matrix elements. FIG.~\ref{fig:proton_2pt} presents behavior of the proton effective mass for ensemble C24P29. Our analysis shows that while the distillation smearing technique alone provides well-optimized plateau regions for the effective mass at low momenta, this stability deteriorates significantly at higher momenta (e.g., $p=1.96\ \mathrm{GeV}$). In comparison, implementation of distillation + momentum smearing scheme yields moderately improved plateau stability at large momenta relative to the distillation-only approach.
\begin{figure*}[pt] 
   \centering
    \includegraphics[width=0.66\textwidth]
    {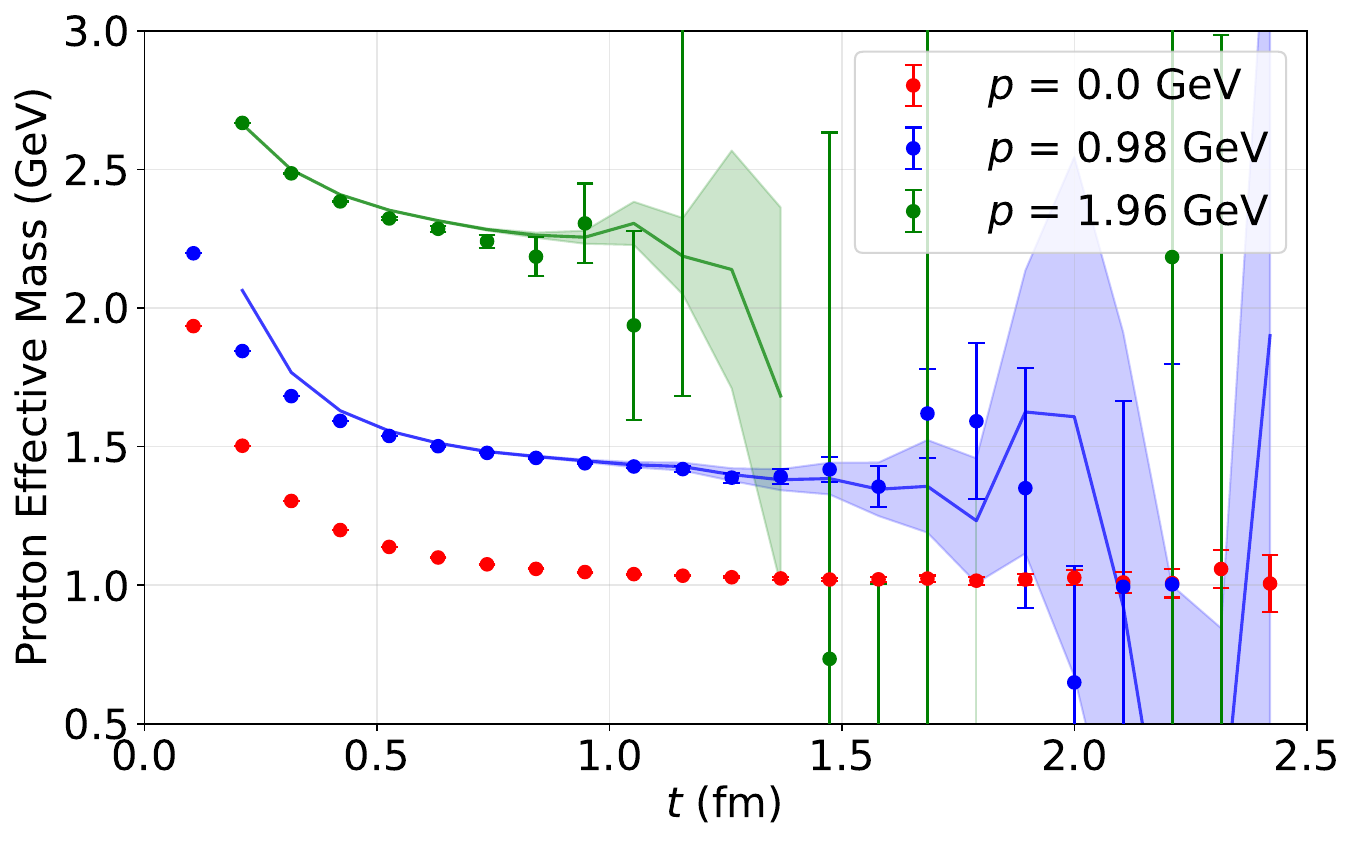} 
   \caption{Proton effective mass with different momentum for ensemble C24P29. The data points with error represent the distillation smearing scheme alone, while the error bands represent the distillation + momentum smearing. The red, green and blue data points (or bands) represent $p=0,0.98,1.96\ \mathrm{GeV}$, respectively.}
\label{fig:proton_2pt}
\end{figure*}
In addition, the three-point correlation function (3pt) can be defined as
\begin{equation}
    \label{eq:3pt_sm}
    C_3(p,t_f,t_i)=\sum_{t_s}\langle 0|\Gamma_i N(p,t_s+t_f)\big(\sum_{\vec{x}}K^{\mu}_{\mathrm{C.G.}}(x,t_i)\big)N^\dagger(p,t_s)|0\rangle,
\end{equation}
with current insertion time slice $t_i$ and $\Gamma_i=\gamma_i\gamma_5$. Among Eq.~\eqref{eq:2pt_sm} and Eq.~\eqref{eq:3pt_sm}, the
topological current bare matrix elements (BMEs) for the momentum $p$ can be obtained through the following ratio
\begin{equation}
    R_{K^{\mu}}(p,t_f,t_i)=\frac{C_3(p,t_f,t_i)}{C_2(p,t_f)},
\end{equation}
which is then fitted to the form
\begin{equation}
\label{eq:ratio_fitting_sm}
    R_{K^{\mu}}(p,t_f,t_i)=\langle K^\mu\rangle_N^{\mathrm{B.}}(p)+c_1e^{-\Delta E(t_f-t_i)}+c_2e^{-\Delta Et_i},
\end{equation}
where $\langle K^\mu\rangle_N^{\mathrm{B.}}$ is defined as nuclear BME divided by the field operator's normalization factor $2E$, and $c_1,c_2,\Delta E$ are fitting parameters.

The matrix elements of the topological current can be effectively discussed by separating its temporal and spatial components. The temporal component $K_t$, dominated by the magnetic field strength term $K_t^{B\cdot A}$, yields matrix elements consistent with zero within uncertainties at zero momentum. These matrix elements then exhibit a gradual increase before saturating to a constant value at higher momenta. Conversely, the spatial components $K_i$, governed primarily by the electric field strength term $K_i^{E\times A}$, maintain approximately constant matrix elements within errors across all momenta studied. This behavior finds a natural interpretation within a point-like proton model, where $\langle K^{\mu}\rangle_{N} \propto S_{\mu}/E$: the charge density (related to $K_t$) is an intrinsic property, while the current density (related to $K_i$) increases from zero at rest toward a finite asymptotic value. These distinct momentum dependencies constitute a direct manifestation of Lorentz covariance in a given frame. Using the same smearing technique employed for the proton effective mass, we extract the BMEs of $K^{t/i}$. The fitted BMEs presented in FIG.~\ref{fig:kmu_BME} validate this theoretical picture, thereby confirming the description outlined above and providing direct evidence for the relations discussed around Eq.~\eqref{eq:tp_ti}.

\begin{figure*}[pt] 
   \centering
   \begin{tabular}{cc}
       \includegraphics[width=0.47\textwidth]{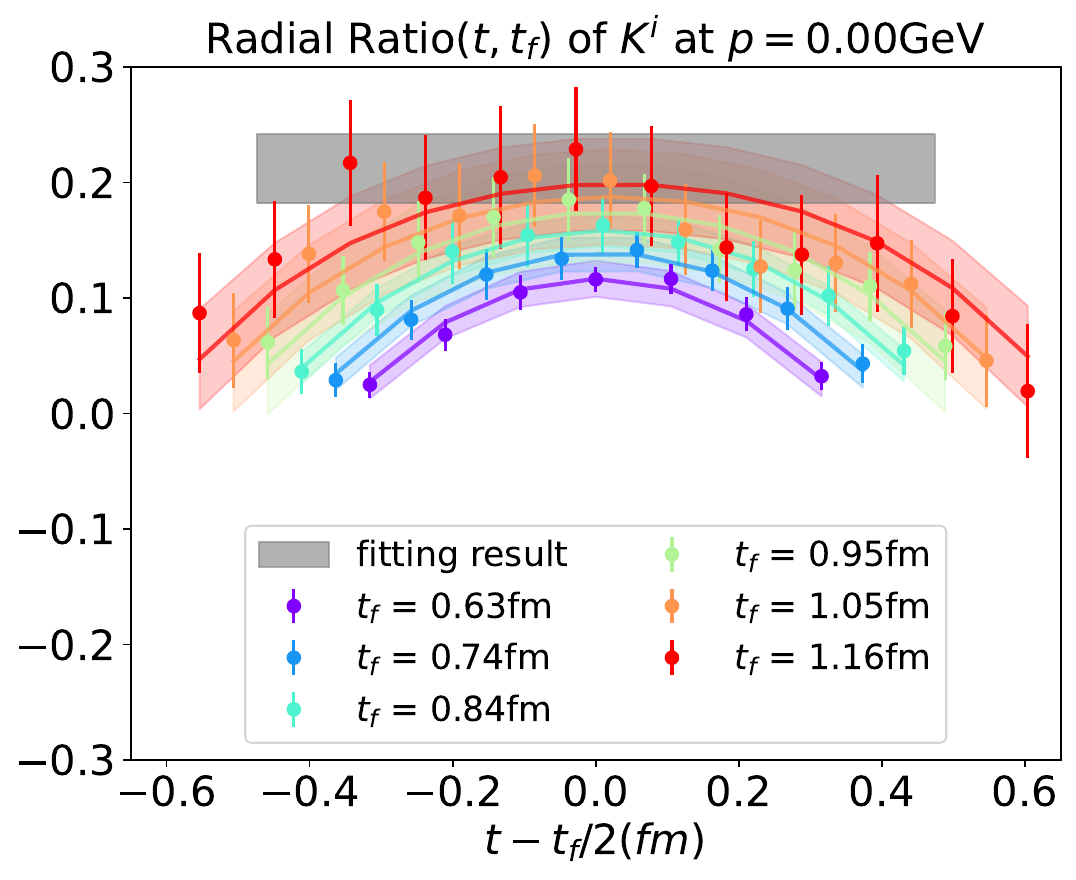}& 
       \includegraphics[width=0.47\textwidth]{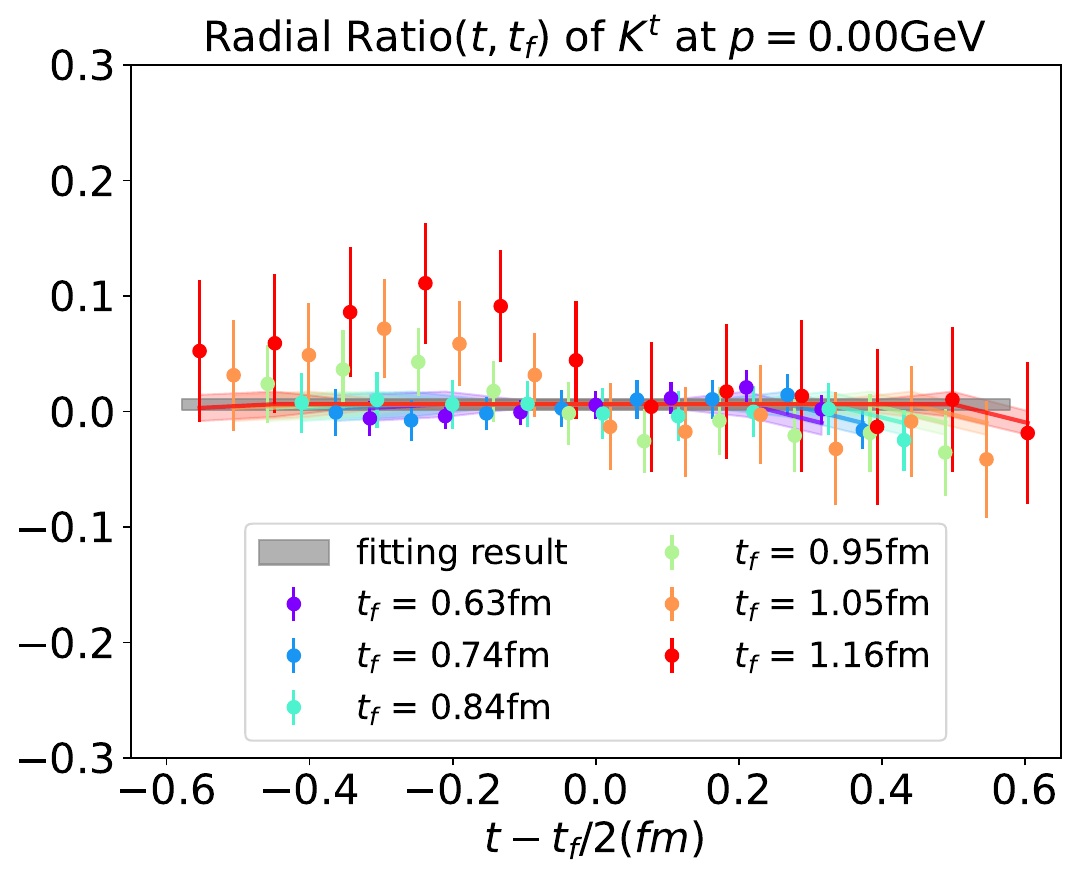} \\
       \includegraphics[width=0.47\textwidth]{figures/BME_fitting/Ki_p2_C24P29.pdf}& 
       \includegraphics[width=0.47\textwidth]{figures/BME_fitting/Kt_p2_C24P29.pdf} \\
       \includegraphics[width=0.47\textwidth]{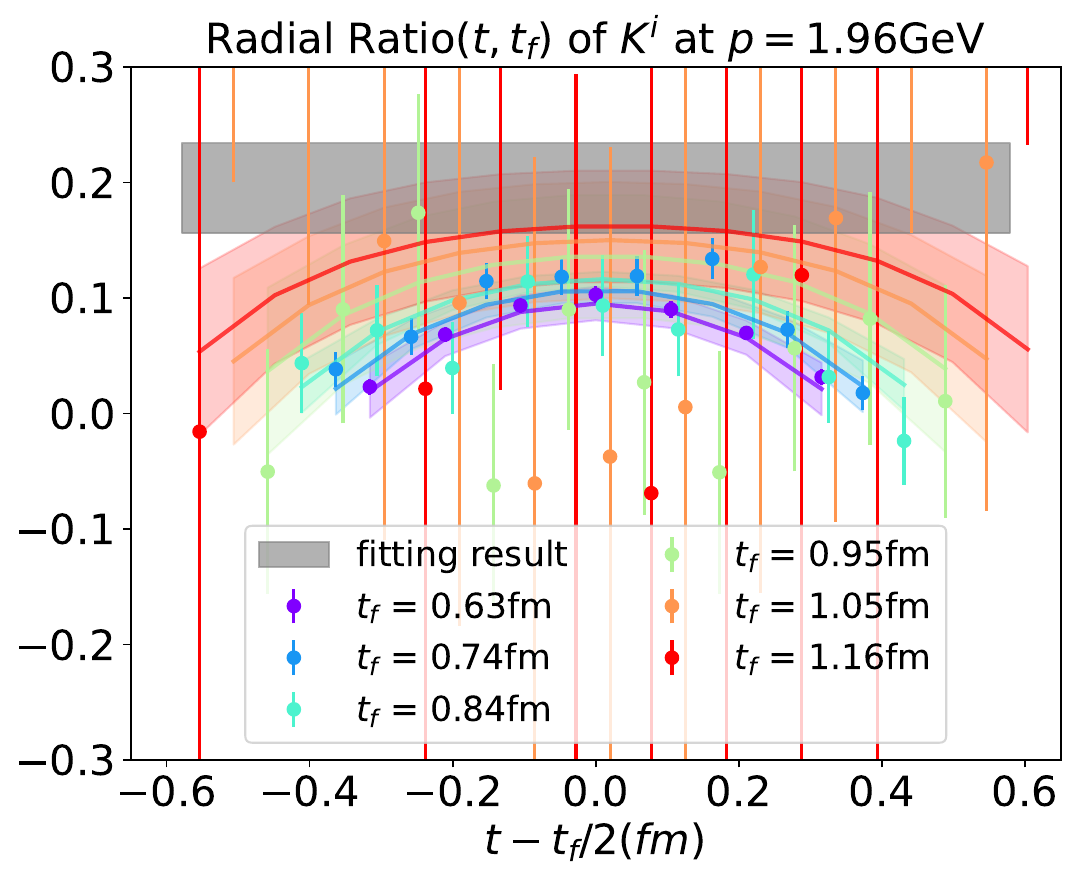}& 
       \includegraphics[width=0.47\textwidth]{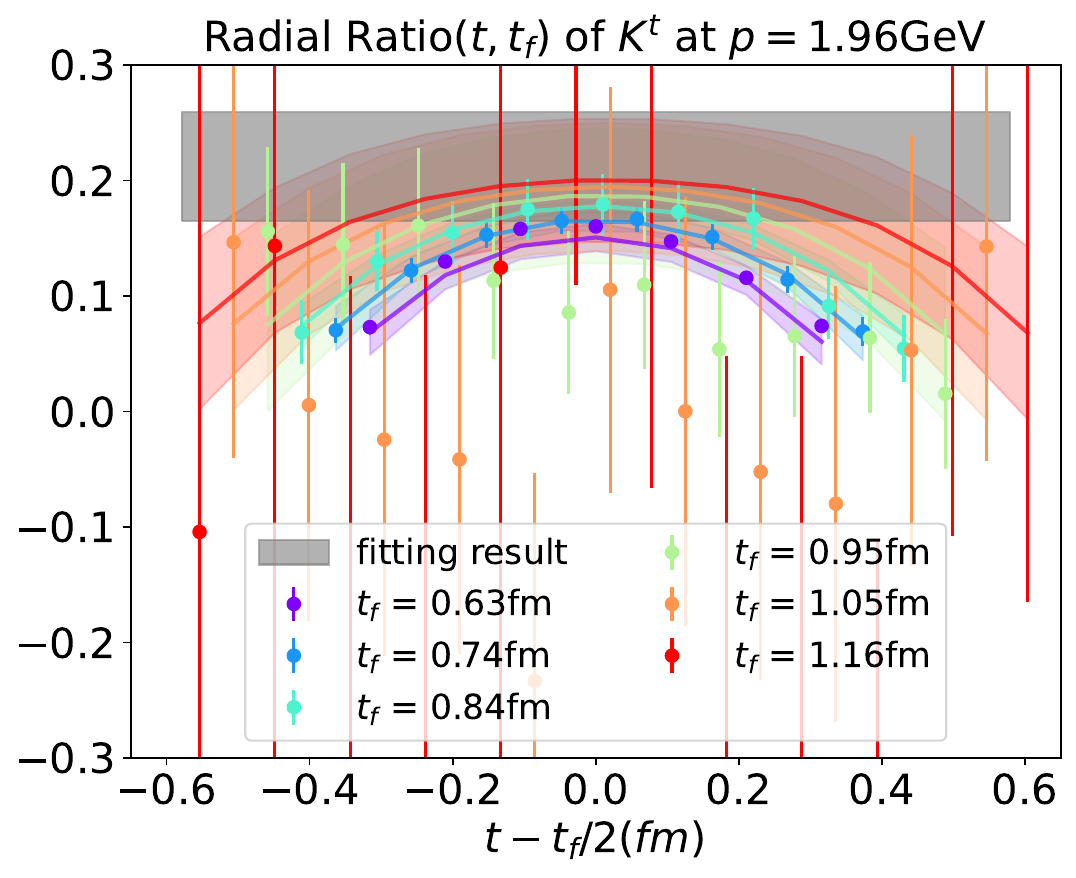} \\
   \end{tabular}
   \caption{Ratio $R_{K^\mu}(t_f, t_i)$ at $t_f\in[0.63,1.16]\ \mathrm{fm}$ and the BMEs $\langle K^\mu\rangle_N^{\mathrm{B.}}$ (the gray error band) for proton external states with varying momenta for ensemble C24P29, utilizing the fitting procedure described in Eq.~\eqref{eq:ratio_fitting_sm}. The results are organized such that the first column corresponds to spatial components $K_i$, while the second column represents temporal component $K_t$. Rows one through three display results for proton momenta of $p=0,0.98,1.96\ \mathrm{GeV}$, respectively.}
    \label{fig:kmu_BME}
\end{figure*}

\subsection{C. Non-perturbative renormalization with cluster decomposition error reduction (CEDR) and $a^2p^2$ correction}\label{sec:renorm}

Within the non-perturbative RI/MOM scheme, the topological current's diagonal renormalization constant $Z_{11}^{\mathrm{RI}}=
K^\mathrm{tree,g}/K^{\mathrm{lat.,g}}$ can be determined as follows
\begin{equation}
    \label{eq:Z11_sm}
    Z_{11}^{\mathrm{RI}}(\mu_{\rm RI}^2)=
    \frac{Z_{g,\mathrm{diag/off}}(\mu_{\rm RI}^2)}{\mathrm{Tr}[S_g^{-1}(p_1)
    G_K(p_1,p_2)
    S_g^{-1}(p_2)(K^{\mathrm{tree,g}})^{-1}]},
\end{equation}
where the external momenta are chosen to satisfy $p_1 = p_2 = p$ and $\mu_{\rm RI}^2=p^2$ denotes the RI/MOM renormalization scale. The bare gluon propagator is given by
\begin{equation}
    \label{eq:gprop_sm}
    S_g^{\mu\nu}(p) =\mathrm{C_V}\langle\mathrm{Tr}[A^\mu(p)A^\nu(-p)]\rangle,
\end{equation}
where $\mathrm{C_V}=2/[(N^2_c-1)\tilde{V}]$, with $N_c=3$ representing the number of colors and $\tilde{V}=\tilde{L}^3\times\tilde{T}$ denoting the dimensionless lattice volume. Due to rotational symmetry breaking on the lattice, we can also define the gluon field renormalization constants through two distinct prescriptions
\begin{eqnarray}
    \label{eq:Zg_sm}
    Z_{g,\mathrm{diag}}^{\mathrm{RI}}(\mu_{\rm RI}^2)&=&\frac{\sum_{p_\rho=0}1}{\hat p^2\sum_{p_\rho=0}S_{g}^{\rho\rho}(p^2)},\nonumber\\
    Z_{g,\mathrm{off}}^{\mathrm{RI}}(\mu_{\rm RI}^2)&=&\frac{\sum_{p_\rho\neq0,p_\tau\neq0,\rho\neq\tau}1}{\hat p^4\sum_{p_\rho\neq0,p_\tau\neq0,\rho\neq\tau}\frac{S_g^{\rho\tau}(p^2)}{\hat p_\rho \hat p_\tau}},
\end{eqnarray}
where $\hat{p}_\mu=\frac{2}{a}\sin{\frac{p_\mu a}{2}}$ represents the lattice momentum. In addition, the computation of $Z_{11}^{\mathrm{RI}}$ necessitates the evaluation of the bare Green's function
\begin{equation}
    \label{eq:bare_green_sm}
    G_K^{\mu\rho\nu}(p_1,p_2)=\sum_{x,y}e^{-i(p_1\cdot x - p_2\cdot y)}\langle A^\mu(x)K^\rho(0)A^\nu(y)\rangle,
\end{equation}
both $A^\mu(x)$ and $K^\mu(x)$ are accessible to direct lattice computation via Eq.~\eqref{eq:amu} and Eq.~\eqref{eq:tp_current_sm}. Ultimately, from all the above formulas, the inverse of $Z_{11}^{\rm RI}$ then reads 
\begin{eqnarray}
    \label{eq:Z11_final_sm}
    Z_{11}^{-1,\mathrm{RI}}(\mu_{\rm RI}^2)&=&\sum_{\mu,\rho,\nu,\sigma}\frac{2\mathrm{C_V}\hat{p}^2
    G_K^{\mu\rho\nu}(p,p)}{i\epsilon^{\rho\mu\sigma\nu}\hat{p}_\sigma(S_g^{\mu\mu}+S_g^{\nu\nu})(p^2)}|_{\mu_{\rm RI}^2=p^2}\nonumber\text{ or}\\
    Z_{11}^{-1,\mathrm{RI}}(\mu_{\rm RI}^2)&=&\sum_{\mu,\rho,\nu,\sigma}\frac{\mathrm{C_V}\hat{p}^4
    G_K^{\mu\rho\nu}(p,p)}
    {i\epsilon^{\rho\mu\sigma\nu}\hat{p}_\sigma\hat{p}_\mu \hat{p}_\nu S_g^{\mu\nu}(p^2)}|_{\mu_{\rm RI}^2=p^2}.
\end{eqnarray}
After matching, the logarithmic divergences are canceled, leaving only a polynomial dependence on the dimensionless RI/MOM scale $a^2p^2$. Therefore, in the subsequent analysis, we employ the relation 
\begin{eqnarray}
    \label{eq:Z11_ms_sm}
    \bar{Z}_{11}^{\mathrm{lat.}}(a^2p^2, \mu^2) = R_{11}(a^2p^2,\mu^2) Z_{11}^{\mathrm{RI}}(a^2p^2),
\end{eqnarray}
and set $\mu^2=10\ \mathrm{GeV}^2$, where $R_{11}$ and $Z_{11}^{\mathrm{RI}}$ are given by Eq.~\eqref{eq:R11exp} and Eq.~\eqref{eq:Z11_final_sm}, respectively. 

Despite achieving a precision better than 1\% in the determination of the gluon propagator (Eq.~\eqref{eq:gprop_sm}), the calculation of the bare gluon Green's function (Eq.~\eqref{eq:bare_green_sm}) remains subject to formidable numerical challenges. 
More concretely, the direct computation of Eq.~\eqref{eq:bare_green_sm} is plagued by substantial statistical uncertainties, making a stable determination of $Z_{11}^{\mathrm{RI}}$ (equivalently $\bar{Z}_{11}^{\mathrm{lat.}}$) at any scale viable only with a statistical sample exceeding hundreds of thousands of configurations.

Applying the cluster decomposition error reduction (CDER) technique enables significant error suppression in $Z_{11}^{\mathrm{RI}}$ (or $\bar{Z}_{11}^{\mathrm{lat.}}$) calculations~\cite{Liu:2017man,Yang:2018bft}. The fundamental principle of cluster decomposition asserts that correlators exhibit exponential decay with increasing separation between operator insertions, implying that integration beyond the correlation length primarily contributes to statistical noise rather than meaningful signal. The CDER methodology addresses this by truncating the volume integral beyond a characteristic distance, thereby achieving an enhancement of the signal-to-noise ratio by a factor of $\sqrt{V}$, as demonstrated in Ref.~\cite{Liu:2017man}. 

The CDER calculation for Eq.~\eqref{eq:bare_green_sm} can be achieved through the following expression
\begin{equation}
    \label{eq:cder_Green}
    G_K^{\mu\rho\nu,\mathrm{CDER}}(p,p)=\langle\int_{|r|<r_1}d^4r \int_{|r^\prime|<r_2}d^4r^\prime\int d^4x e^{ip\cdot r^\prime}K^\rho(x+r^\prime) \mathrm{Tr}[A_\mu(x)A_\nu(x+r)]\rangle,
\end{equation}
where we introduce two distinct cutoff parameters: $r_1$ between the gluon operator and one gauge field, and $r_2$ between the gauge fields within the gluon propagator. This implementation yields the modified correlator with appropriate spatial cutoffs. The CDER technique achieves computational efficiency of $\mathcal{O}(V\log V)$ through the repeated application of fast Fourier transforms (FFTs). 
In concrete terms, Eq.~\eqref{eq:cder_Green} can be realized through the strategy outlined below:

1) Construct the spatially-restricted topological current operator: We compute $K^{\rho}_{r^\prime}(x)=\int_{|r^\prime|<r_2}d^4r^\prime K^\rho(x+r^\prime)$ by first performing Fourier transforms (FT) of both $K^\rho(x)$ and the Heaviside function $f(x)=\theta(r_2-|x|)$, multiplying the transformed functions in momentum space, and subsequently applying anti-FT.

2) Form the intermediate operator product: We calculate $B^{\rho}_{\mu,r^\prime}(x)=A_\mu(x)K^{\rho}_{r^\prime}(x)$ through direct multiplication in coordinate space.

3) Apply cluster decomposition to the correlation function: For the expression $\int d^4xd^4ye^{ip\cdot(x-y)}B^{\rho}_{\mu,r^\prime}(x)A_\nu(y)$, we perform FT of both $A_\nu$ and $B^{\rho}_{\mu,r^\prime}$, apply anti-FT to the product $A_\nu(p)B^{\rho}_{\mu,r^\prime}(-p)$, implement an additional Heaviside function $g(x)=\theta(r_1-|x|)$ in coordinate space, and FT the resulting product.

The selection of cutoff parameters $r_{1,2}$ in the CDER technique requires careful consideration, as inappropriate choices may induce periodic oscillations in the renormalization constants~\cite{Liu:2017man}. FIG.~\ref{fig:CDER_cut} provides visual confirmation of this analysis under ensemble C48P23. When the cutoff parameter governing the gluon field-operator correlation ($r_1$) is set too small (top left panel), distinct periodic oscillations become apparent in the results. These oscillations progressively stabilize as $r_1$ increases. A similar pattern is observed for variations in the current cutoff parameter ($r_2$), though changes in oscillatory behavior are less pronounced without corresponding adjustments to $r_1$. Based on the above discussion, we employ the optimal cutoffs identified in the bottom right panel of FIG.~\ref{fig:CDER_cut}—specifically, $r_1 = 1.26\ \mathrm{fm}$ and $r_2 = 0.74\ \mathrm{fm}$—to perform the subtraction of discretization errors. This choice of cutoff parameters aligns with the observation in Ref.~\cite{Yang:2018bft} that $r_1 \approx 1.4 r_2$ yields the optimal signal-to-noise performance. 

\begin{figure*}[pt] 
   \centering
   \begin{tabular}{cc}
       \includegraphics[width=0.47\textwidth]{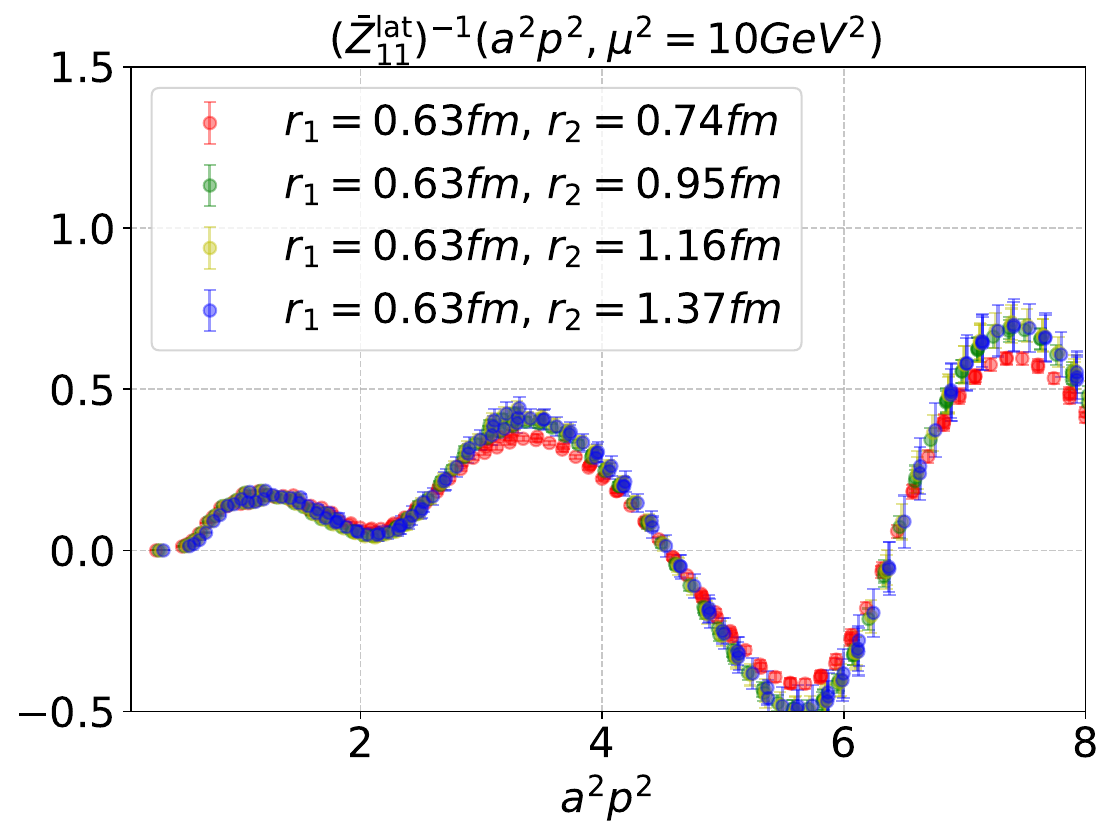}& 
       \includegraphics[width=0.47\textwidth]{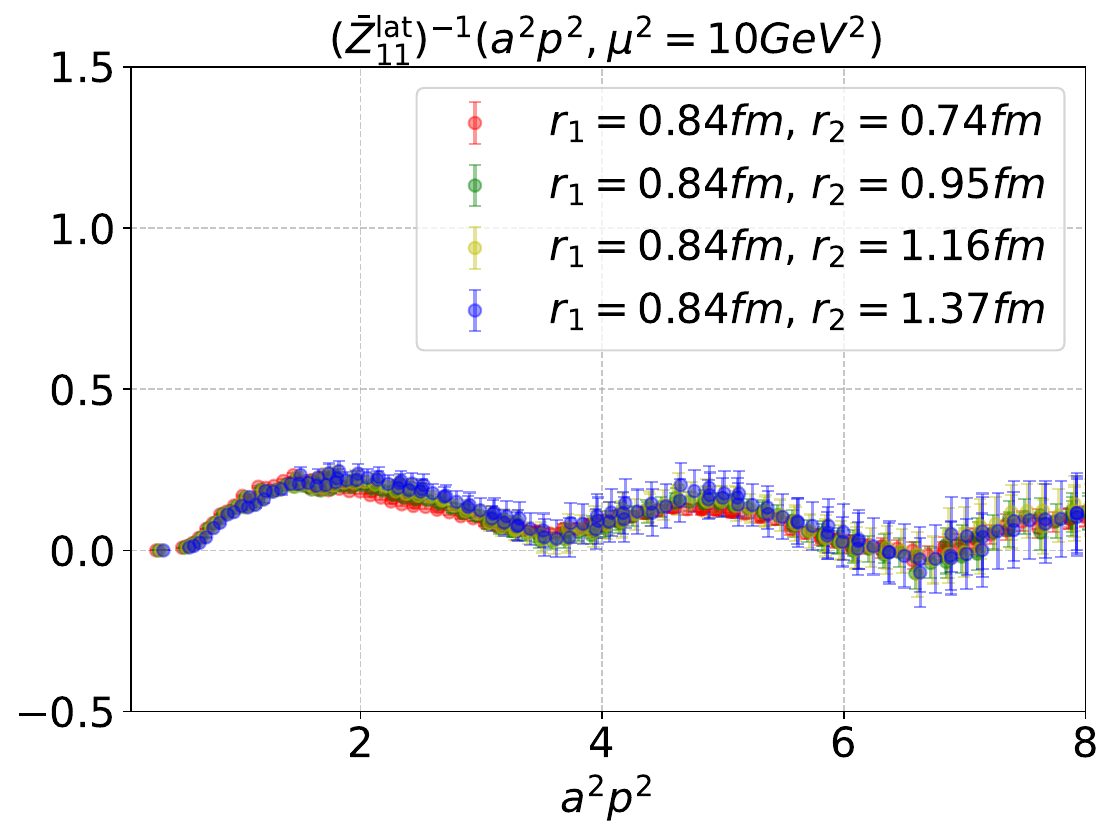} \\
       \includegraphics[width=0.47\textwidth]{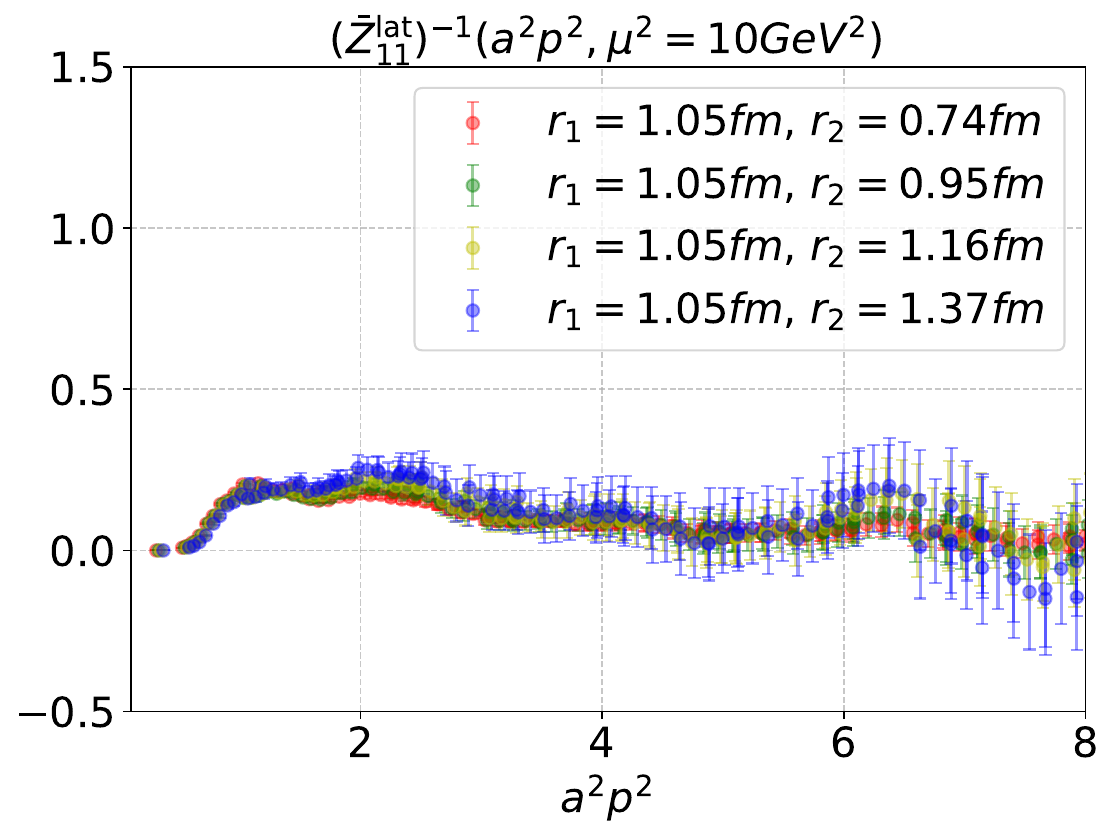}& 
       \includegraphics[width=0.47\textwidth]{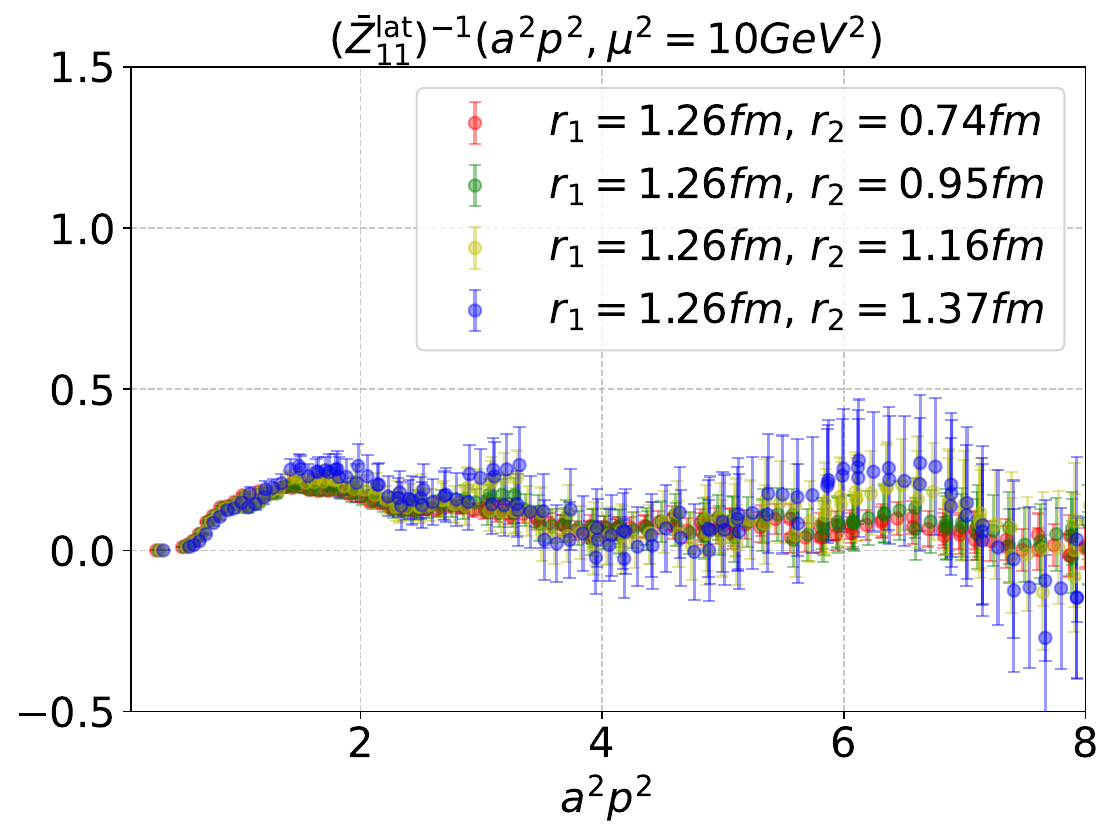} \\ 
   \end{tabular}
   \caption{Performance of the renormalization constant $\bar{Z}_{11}^{\mathrm{lat.}}(a^2p^2, {\mu}^2=10\ \mathrm{GeV}^2)$ for various cutoff parameters $r_{1,2}$ under ensemble C48P23 using momentum modes $(p,p,p,p)$.}
    \label{fig:CDER_cut}
\end{figure*}

As established, $\bar{Z}_{11}^{\mathrm{lat}.}$ carries a discretization error that depends on $a^2 p^2$. However, as evidenced by Fig.~\ref{fig:CDER_cut}, within the effective window of $a^2 p^2$—determined by considering both perturbative matching and the signal-to-noise ratio of the lattice data—this discretization error remains too large to permit a reliable fit. Consequently, we employ the method from Ref.~\cite{Yang:2018nqn} to suppress the $a^2 p^2$ corrections. Specifically, for the 5-HYP smeared topological current, most of the $a^2 p^2$ dependence in $\bar{Z}{11}^{\mathrm{lat}.}$ can be eliminated by constructing the following ratio:
\begin{equation}
    \label{eq:sub_Z11}
    \tilde{\bar{Z}}_{11}^{\mathrm{lat.}}(a^2p^2,\mu^2)=\bar{Z}_{11}^{\mathrm{lat.}}(a^2p^2,\mu^2)f(a^2p^2\rightarrow0)/f(a^2p^2),\quad f(a^2p^2)=\frac{\langle \mathrm{Tr}[A_\mu^{5\mathrm{-HYP}}(p)A_\mu^{5\mathrm{-HYP}}(-p)]\rangle}{\langle \mathrm{Tr}[A_\mu(p)A_\mu(-p)]\rangle},
\end{equation}
where $A_\mu^{5\mathrm{-HYP}}$ is the 5-HYP-smeared gauge potential defined from the 5-HYP-smeared gauge link, and $f(a^2p^2)$ is the smeared ratio of two gluon propagators (Eq.\eqref{eq:gprop_sm}) with and without HYP smearing. Fig.~\ref{fig:sub_Z11} demonstrates the significance of the $a^2 p^2$ correction. Prior to correction, the renormalization constant exhibits a severe dependence on $a^2 p^2$, rendering the subtraction of discretization errors unpredictable. After applying the correction, although the signal-to-noise ratio deteriorates—primarily because the factor $f(a^2p^2\rightarrow0)/f(a^2p^2)$ itself becomes large at high $a^2 p^2$—the result forms a plateau near unity. This behavior aligns with the physical expectation for a renormalization constant. 

\begin{figure*}[pt] 
   \centering
    \includegraphics[width=0.66\textwidth]
    {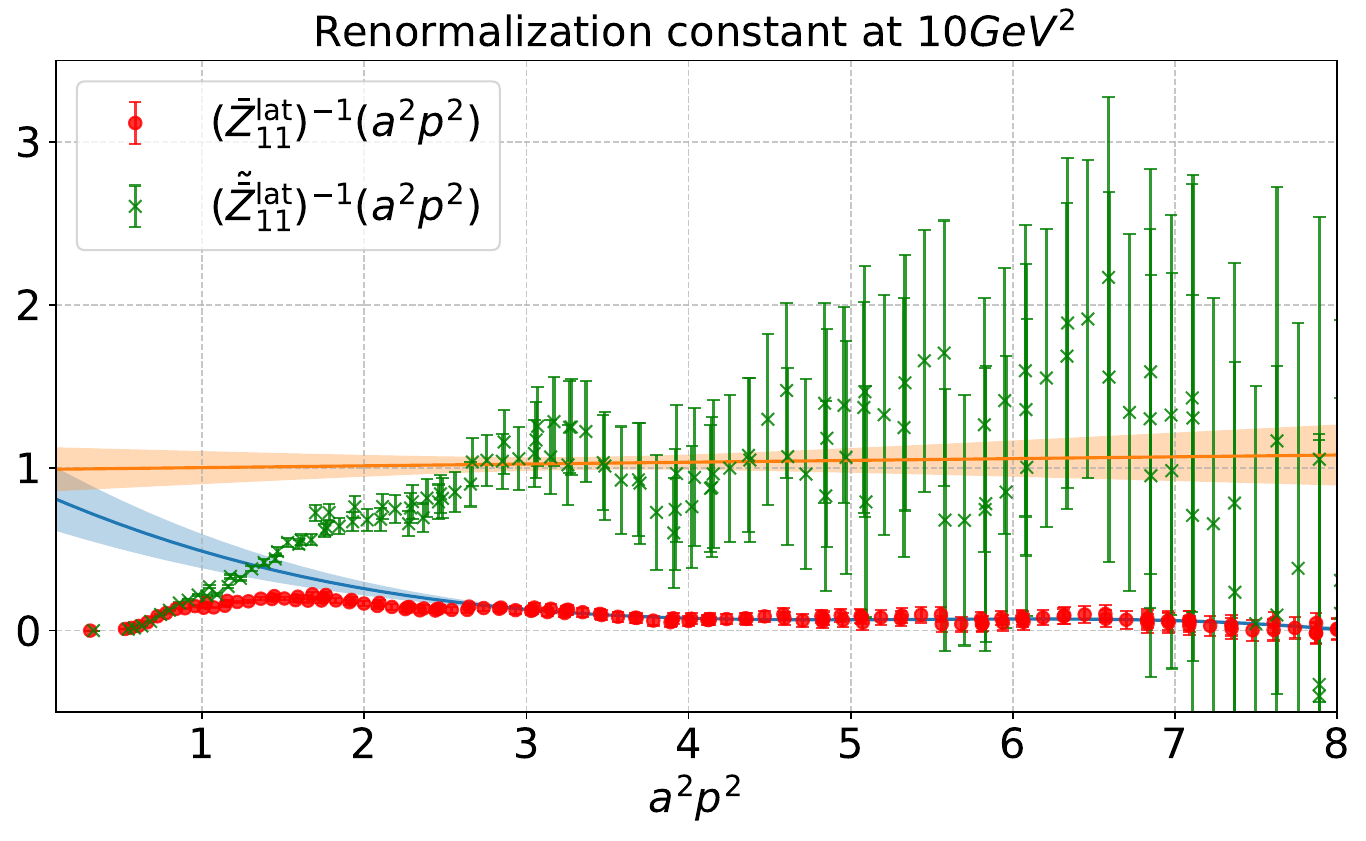} 
   \caption{Behavior of the renormalization constant under three-loop matching for ensemble C48P23, comparing results with (green) and without (red) the $a^2p^2$ correction from Eq.~\eqref{eq:sub_Z11}. The corresponding error bands represent fits that subtract discretization errors using an $a^2p^2$ polynomial: a third-order polynomial for the green band and a first-order polynomial for the red band. The fitted interval is $a^2p^2\in[2.6, 7.1]$.}
\label{fig:sub_Z11}
\end{figure*}

The above analysis based for other ensembles (such as E32P29 and F32P30 in Table~\ref{tab:ensem_total}) give similar results. Thus, we have successfully determined the purely gluonic renormalization constant by employing the CDER technique combined with $a^2 p^2$ correction. The CDER technique effectively suppresses the severe statistical noise inherent in direct calculations by truncating the volume integral beyond characteristic distances $r_1$ and $r_2$. Concurrently, the $a^2 p^2$ correction significantly reduces the discretization errors in the renormalization constant, yielding reliable physical results. Together, these techniques allow us to determine this quantity with acceptable error—a task previously feasible only by accumulating enormous statistics. 

Building upon all the preceding analysis, we perform a combined extrapolation to the continuum limit and the IMF, namely
\begin{equation}
    \label{eq:global_fit_sm}
    \langle K^\mu\rangle_{N}=\bar{Z}_{11}^{\mathrm{lat.}}\langle K^\mu\rangle_N^{\mathrm{B.}}=\frac{S^\mu}{E}\big(\Delta G+\frac{c_{\mathrm{h.t.}}}{E^2}\big) + c_a a^2,
\end{equation}
where $c_{\mathrm{h.t.}}$ accounts for higher-twist contributions, and $c_a$ captures discretization effects. 
Since the proton momenta are not significantly larger than its mass, we incorporate the higher-twist contribution using an empirical parametrization form $1/E^2$ in the equation above, where $E = \sqrt{M^2 + p^2}$ with the proton mass $M\approx1\ \mathrm{GeV}$. To maintain clarity and avoid symbol redundancy, we adopt the notation $\bar{Z}_{11}^{\mathrm{lat}.}$ in place of $\tilde{\bar{Z}}_{11}^{\mathrm{lat}.}$ in the following discussion. This extrapolation utilizes data from three distinct lattice spacings and five proton momentum values, with the input raw data provided in Table~\ref{tab:final_result}. In addition, the primary sources of systematic uncertainty in our analysis are identified as follows: 

1) Including or not the zero-momentum data points in the extrapolation;

2) Discretization errors associated with the finite lattice spacing: the central value difference between the zero momentum $\langle K_i\rangle_N$ on finest lattice spacing and that extrapolated to the continuum limit; 

3) The choice of the UV window in the RI/MOM scale or $a^2p^2$: the interval $\mu_{\mathrm{RI}}\in[3.0, 5.0]\ \mathrm{GeV}$ has been extended to $[3.0, 6.0]\ \mathrm{GeV}$ to estimate the corresponding uncertainty;

4) Systematic errors associated with the perturbative inputs: estimated by varying from three-loop to two-loop accuracy. 

As illustrated in FIG.~\ref{fig:continue_extra_sm}, the joint extrapolation yields a gluon helicity contribution $\Delta G^{\overline{\mathrm{MS}},\ 10\ \mathrm{GeV}^2} = 0.231(17)^{\mathrm{sta.}}(44)^{\mathrm{sym.}}$, which constitutes $46(9)\%$ of the proton's total spin. 
\begin{table*}
\resizebox{0.66\columnwidth}{!}{
\begin{tabular}{|c|c|c|c|c|c|}
\hline
\multirow{2}{*}{$a$ (fm)} & \multicolumn{3}{c|}{$\langle K^\mu\rangle_N^{\mathrm{B.}}$} & \multicolumn{2}{c|}{$\bar{Z}_{11}^{\mathrm{lat}.}$} \\ 
\cline{2-6}
 & $p(\mathrm{GeV})$ & $\langle K_t\rangle_N^{\mathrm{B.}}$ & $\langle K_i\rangle_N^{\mathrm{B.}}$ & $2\mathrm{-loop}$ & $3\mathrm{-loop}$\\
\hline
\multirow{6}{*}{0.10524(05)(62)} & 0.00 & 0.006(05) & 0.212(30) & \multirow{6}{*}{0.91(09)(03)} & \multirow{6}{*}{1.09(12)(04)} \\
 & 0.49 & 0.110(18) & 0.209(11) & & \\
 & 0.98 & 0.154(25) & 0.187(32) & & \\
 & 1.47 & 0.195(38) & 0.180(34) & & \\
 & 1.96 & 0.212(43) & 0.202(46) & & \\
 & 2.45 & 0.214(46) & 0.152(93) & & \\
\hline
\multirow{3}{*}{0.08973(20)(53)} & 0.00 & 0.004(07) & 0.164(07) & \multirow{3}{*}{0.89(10)(04)} & \multirow{3}{*}{1.07(12)(04)} \\
 & 0.86 & 0.193(16) & 0.202(53) & & \\
 & 1.29 & 0.214(53) & 0.171(63) & & \\
\hline
\multirow{3}{*}{0.07753(03)(45)} & 0.00 & 0.001(07) & 0.166(09) & \multirow{3}{*}{0.85(05)(03)} & \multirow{3}{*}{0.98(06)(03)} \\
 & 1.00 & 0.181(10) & 0.165(12) & & \\
 & 1.50 & 0.173(30) & 0.176(32) & & \\
\hline
\end{tabular}}
\centering
\caption{Raw data employed for the global fitting corresponding to Eq.~\eqref{eq:global_fit_sm}. The total uncertainty in the renormalization constants (for both the two- and three-loop matching) comprises two components: the statistical error and systematic error, which the systematic error given by difference between the center value of RG-evolved result and fixed-order result.}
\label{tab:final_result}
\end{table*}

\begin{figure*}[pt] 
   \centering
    \includegraphics[width=0.66\textwidth]
    {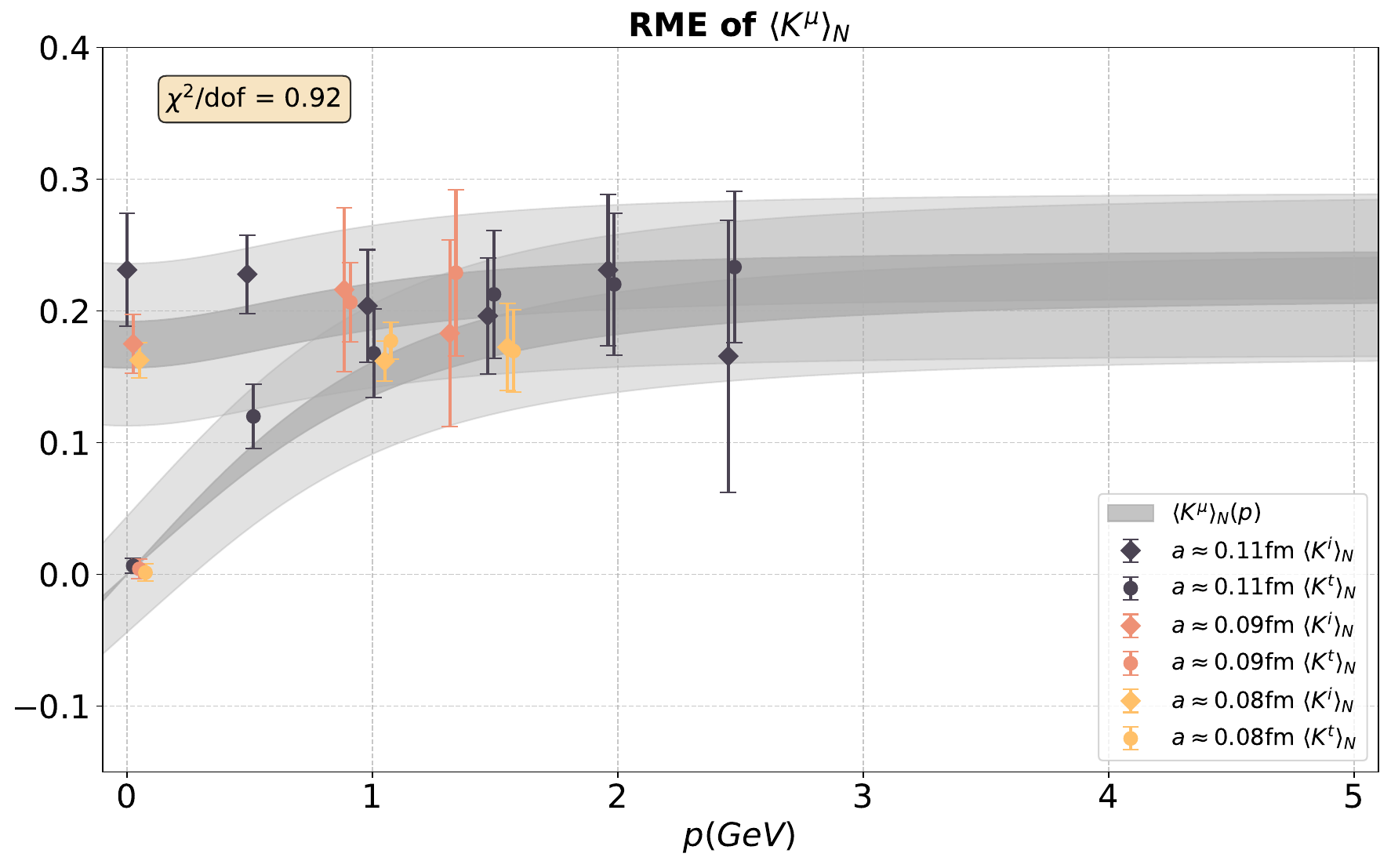}
   \caption{The continuum and infinite momentum extrapolation of the total gluon helicity is performed using lattice spacings from $0.08\ \mathrm{fm}$ to $0.11\ \mathrm{fm}$ at the $\overline{\mathrm{MS}}$ scale $\mu^2=10\ \mathrm{GeV}^2$. The extrapolation employs the functional form given in Eq.~\eqref{eq:global_fit_sm}, with the raw input data provided in Table~\ref{tab:final_result}. The uncertainties are shown as separate bands: statistical uncertainty (dark gray) and all uncertainties that arise during our analysis (light gray).}
\label{fig:continue_extra_sm}
\end{figure*}

\end{widetext}

\end{document}